\documentclass{sig-alternate}

\usepackage{url}
\usepackage{amsmath}
\usepackage{amssymb}
\usepackage{bm}
\usepackage{stmaryrd}
\usepackage{color}
\usepackage{colortbl}
\usepackage{epsf}

\newcommand{\ignore}[1]{}

\ignore{
\newtheorem{proposition}{Proposition}
\newtheorem{lemma}{Lemma}
\newtheorem{corollary}{Corollary}
\newtheorem{theorem}{Theorem}
\newtheorem{definition}{Definition}

\newtheorem{example}{Example}
}

\DeclareMathAlphabet{\mathpzc}{OT1}{pzc}{m}{it}

\def\be{\begin{equation}}
\def\ee{\end{equation}}
\def\bea{\begin{eqnarray}}
\def\eea{\end{eqnarray}}
\def\nn{\nonumber}

\def\s{\sigma}

\def\l{\left}
\def\r{\right}

\def\ss{\smallskip}

\def\w{\wedge}
\def\ra{\rightarrow}

\def\P{\Phi}

\def\fa{\forall}

\def\bc{\bigcap}
\def\S{\Sigma}
\def\nin{\not\in}

\def\ex{\exists}

\def\hb{\hbox}
\def\lar{\leftarrow}

\def\sc{\scriptsize}

\def\rl{\rightleftharpoons}

\def\ba{\begin{align}}
\def\ea{\end{align}}
\def\bes{\begin{split}}
\def\es{\end{split}}
\def\vD{\vDash}

\newcommand{\comlb}[1]{{\vspace{2mm}\noindent \bf COMM(LEO):}~ #1 \hfill {\bf
    END.}\\}
\newcommand{\comj}[1]{{\vspace{2mm}\noindent \bf COMM(JAF):}~ #1 \hfill {\bf
    END.}\\}

\newcommand{\defproof}[2]{{\noindent\bf Proof of #1:\
}#2 \boxtheorem}

\newcommand{\boxtheorem}{\hfill $\Box$}
\newcommand{\nit}[1]{{\it #1}}

\newcommand{\mdg}{\nit{MDG}(M)}
\newcommand{\mc}[1]{\mathcal{ #1}}

\begin{document}

\title{Matching Dependencies with Arbitrary Attribute Values: Semantics, Query Answering and Integrity Constraints\titlenote{Research supported by the NSERC Strategic Network on Business Intelligence (BIN,ADC05) and NSERC/IBM CRDPJ/371084-2008.} }
\numberofauthors{3}
\author{
\alignauthor Jaffer Gardezi\\
\affaddr{University of Ottawa, SITE}\\
\affaddr{Ottawa, Canada}\\
\email{jgard082@uottawa.ca}
\alignauthor Leopoldo Bertossi\titlenote{Faculty Fellow of the IBM CAS. Also affiliated to University of
Concepci\'{o}n (Chile).}\\
\affaddr{Carleton University, SCS}\\
\affaddr{Ottawa, Canada}\\
\email{bertossi@scs.carleton.ca}
\alignauthor Iluju Kiringa\\
\affaddr{University of Ottawa, SITE}\\
\affaddr{Ottawa, Canada}\\
\email{kiringa@site.uottawa.ca}
}

\maketitle
\begin{abstract}
Matching dependencies (MDs) were introduced to specify the identification or matching of certain attribute values
in pairs of database tuples when some similarity conditions are satisfied. Their enforcement can be seen as
a natural generalization of entity resolution. In what we call the {\em pure case} of MDs, any value from the underlying
data domain can be used for the value in common that does the matching.
We investigate the semantics and properties of data cleaning through the enforcement of matching dependencies for the
pure case. We characterize the intended clean instances and also the {\em clean answers} to queries as those that are invariant under the cleaning process. The complexity of computing clean instances and clean answers to queries
is investigated. Tractable and intractable cases depending on the MDs and queries are identified. Finally,
we establish connections with database {\em repairs} under integrity constraints.
\end{abstract}




\ignore{\newtheorem{theorem}{Theorem}
\newtheorem{lemma}{Lemma}
\newtheorem{corollary}{Corollary}
\newtheorem{proposition}{Proposition}}
\newdef{theorem}{Theorem}
\newdef{lemma}{Lemma}
\newdef{corollary}{Corollary}
\newdef{proposition}{Proposition}
\newdef{definition}{Definition}
\newdef{example}{Example}

\section{Introduction}

A database instance may contain several tuples and values in them that refer to the same external
entity that is being modeled through the database. In consequence, the database may be
  modeling the same entity in different forms, as different entities, which most likely is not the intended representation.
This problem  could be caused by errors in data, by data coming from  different sources that use different formats or semantics,
  etc. In this case, the database is considered to contain dirty data, and it must undergo a cleansing process that goes through two interlinked phases: detecting tuples (or values therein) that should be
matched  or identified, and, of course, doing the actual matching. This problem is usually called {\em entity
resolution}, {\em data fusion}, {\em duplicate record detection}, etc. Cf. \cite{elmargamid,naumannACMCS} for some recent surveys and \cite{BenjellounGMSWW09} for recent work in this area.

Quite recently, and generalizing entity resolution, \cite{Fan08,Fan09} introduced {\em matching dependencies} (MDs), which are declarative specifications of matchings of attribute values that should hold under certain conditions.
MDs help  identify duplicate data and enforce their merging
by exploiting semantic knowledge expressed.

Loosely speaking,
an MD is a rule defined on a database
which states that, for any pair of tuples from given relations
within the database, if the values of certain attributes of the tuples are
similar, then the values of another set of attributes should be considered to
represent the same object. In consequence, they should take the same values. Here, similarity
of values can mean equality or a domain-dependent similarity relationship, e.g. related to some metric,
such as the edit distance.

 \begin{example}
\label{ex:md}
Consider the following database instance of a relation $P$.

\ss
\begin{tabular}{|c|c|c|}
Name & Phone & Address \\ \hline
John Smith & 723-9583 & 10-43 Oak St. \\
J. Smith & (750) 723-9583 & 43 Oak St. Ap. 10  \\
\end{tabular}
\ss

\noindent Similarity of the names in the two tuples
(as measured by, e.g. edit distance) is insufficient to
establish that the tuples \ refer to the \ same person. This is
because the last name is a common one, and only the first
initial of one of the names is given. However, similarity of
their phone and address values indicates that
the two tuples may be duplicates. This is expressed by
an MD which states that, if two tuples from $P$
have similar address and phone, then the names
should match. In the notation of MDs, this is expressed as

\vspace{2mm}
$P[Phone]\approx P[Phone]\w P[Address]\approx P[Address] \ \ra$\\
\hspace*{3.9cm}$P[Name]\rl P[Name]$.
\boxtheorem
\end{example}
The identification in \cite{Fan08,Fan09} of a new class of dependencies and their declarative formulation have become important additions to  data cleaning research. In this work we investigate matching dependencies, starting from and refining the model-theoretic and dynamic semantics of MDs introduced in \cite{Fan09}.

 Any method of querying a dirty data source must address the issue of duplicate detection in order to
obtain accurate answers. Typically, this is done by first cleaning the data
by discarding or combining duplicate tuples and standardizing formats. The result will be a new database
where the entity conflicts have been resolved. However, the entity resolution problem may have different {\em solution instances} (which we will simply call {\em solutions}), i.e. different clean
versions of the original database. The model-theoretic semantics that we propose and investigate defines and
 characterizes the class of solutions, i.e. of intended clean instances.

After a clean instance has been obtained, it  can be queried as usual.
  However, the query answers will then depend on the particular solution at hand. So, it becomes
relevant to characterize those query answers that are invariant under the different (sensible) ways of
cleaning the data, i.e. that persist across the solutions. This is an interesting problem {\em per se}. However, it becomes crucial if one wants to obtain semantically
clean answers while still querying the original dirty data source.

This kind of virtual cleaning and query answering on top of it have been investigated
in the area of {\em consistent query answering} (CQA) \cite{Arenas99}, where, instead of MDs, classical integrity constraints (ICs) are
considered, and database instances are {\em repaired} in order to restore consistency (cf. \cite{bc-2003,B2006,chom07} for
surveys of CQA). Virtual approaches to robust query answering under entity resolution and enforcement of
matching dependencies are certainly
unavoidable in virtual data integration systems.

In this paper we make the following contributions, among others:
\begin{enumerate}
\item We revisit the semantics of MDs introduced in \cite{Fan09}, pointing out sensible
and justified modifications of it.
A new semantics for MD satisfaction is then proposed and formally developed.

\item Using the new MD semantics, we formally define the intended solutions
for a given, initial instance, $D_0$, that may not satisfy a given set of MDs. They are called
{\em minimally resolved instances} (MRIs) and are obtained through an iteration process that
stepwise enforces the satisfaction of MDs until a stable instance is reached. The resulting
instances minimally  differ from $D_0$ in terms of number of changes of attribute values.

This semantics (and the whole paper) considers the {\em pure case} introduced in \cite{Fan09}, in the sense that
the values than can be chosen to match attribute values are arbitrarily taken from the
underlying data domains. No matching functions are considered, like in \cite{BenjellounGMSWW09}, for example (where entire tuples are merged, not individual attribute values).

\item We introduce the notion of {\em resolved answers} to a query posed to $D_0$. They are the answers
that are invariant under the MRIs.

\item We investigate the computability and complexity of computing MRIs and resolved answers, identifying syntactic conditions on MDs and conjunctive queries
under which the latter becomes tractable via query rewriting. The rewritten queries are allowed to contain
counting and transitive closure (recursion).

\item We identify cases where computing (actually, deciding) resolved answers is coNP-complete.

\item We establish a connection between MRIs and database repairs under functional
dependencies as found in CQA. In the latter case, the repairs consider, as usual, a notion of minimality based on deletion of
whole tuples and comparison under set inclusion. This reduction allows us to profit from results
in CQA, obtaining additional (in)tractability results for MDs.
\end{enumerate}
This paper is organized as follows. Section \ref{sec:pre} presents basic concepts and
notations needed in the rest of the paper.  Section \ref{sec:prob}
identifies some problems with the MD semantics, and refines it
to address them. It also
introduces the resolved instances and resolved answers to
a query. Section \ref{subsec:com} considers the problems of computing resolved instances and
resolved query answers.  Section
\ref{sec:Tr} identifies queries and sets
of MDs for which computing resolved
answers becomes tractable via query rewriting.
Section \ref{subsec:NI2} establishes the connection with CQA.
 Section \ref{sec:con} presents some final
conclusions.

\section{Preliminaries}
\label{sec:pre}

In general terms, we consider a relational schema $\mathcal{S}$ that includes an enumerable infinite domain $U$. An instance
$D$ of $\mathcal{S}$ can be seen as a finite set of ground atoms of the form $R(\bar{t})$, where $R$ is a database predicate
in $\mathcal{S}$, and $\bar{t}$ is a tuple of constants from $U$. We assume that each database tuple has an identifier,
e.g. an extra attribute that acts as a key for the relation and is not subject to updates. In the following it will not
be listed, unless necessary, as one of the attributes of a database predicate. It plays an auxiliary role
only, to keep track of updates on the other attributes. $R(D)$ denotes the extension
of $R$ in $D$. We sometimes refer to attribute $A$ of  $R$ by $R[A]$.
If the $i$th attribute of predicate $R$ is $A$, for a tuple $t = (c_1, \ldots, c_j) \in R(D)$,
$t[A]$ denotes the value $c_i$. The symbol $t[\bar A]$ denotes the vector whose entries are the
values of the attributes in the vector $\bar A$. The attributes may have subdomains that are contained
in $U$. Constants will be denoted by lower case letters at the beginning of the alphabet\ignore{, and
distinct letters will denote unequal constants}.

A matching dependency \cite{Fan08}, involving predicates\linebreak $R(A_1,\ldots,A_n)$, $S(B_1,\ldots,B_m)$, is
 a rule of the form
{\small \bea \label{eq:md}
\bigwedge_{i \in I, j \in J}R[A_i] \approx_{ij} S[B_j]  \ra \bigwedge_{i \in I', j \in J'}R[A_i]\rl S[B_j].
\eea}
Here $R$ and $S$ could be the same predicate. $I, I'$ and $J,J'$ are fixed subsets of $\{1, \ldots, n\}$ and
$\{1, \ldots, m\}$, resp. We assume that, when $A_i, B_j$ are related via $\approx_{ij}$ or $\rl$ in (\ref{eq:md}),
they share the same (sub)domain, so their values can be
compared by the domain-dependent binary  similarity predicate, $\approx_{ij}$ or can be identified, resp.

The similarity operators, generically denoted with $\approx$, are assumed to have the properties of: (a) Symmetry: If $x\approx y$, then $y\approx x$.
(b) Equality subsumption: If $x = y$, then $x\approx y$.

 The MD in (\ref{eq:md}) is implicitly universally quantified in front and applied to pairs of tuples $t_1, t_2$ for $R$ and $S$, resp.
The expression $\bigwedge R[A_i]\approx_{ij} S[B_j]$   states that the values
of the attributes $A_i$ in tuple $t_1$ are similar to those of attributes $B_j$ in
tuple $t_2$. If this holds,
 the expression $R[A_i]\rl S[B_j]$ indicates that, for the same tuples
$t_1$ and $t_2$,  $t_1[A_i]$ and
$t_2[B_j]$ on the RHS should be updated so that they become the same, i.e. their  values are identified
or matched.
However, the attribute values to be used for this matching are left unspecified by (\ref{eq:md}).

For abbreviation, we will sometimes write MDs as
\bea
R[{\bar A}] \approx S[{\bar B}]\ra R[\bar C]\rl S[\bar E], \label{eq:simplMD}
\eea
where $\bar A$, $\bar B$, $\bar C$, and $\bar D$ represent the lists of attributes,
$(A_1, ...,A_k)$, $(B_1, ...,B_k)$, $(C_1, ..., C_{k'})$, and $(E_1, ...,E_{k'})$, respectively.
We refer to the pairs of attributes $(A_i,B_i)$ and $(C_i,E_i)$
as {\em corresponding pairs} of attributes of the pairs $({\bar A}, {\bar B})$ and $({\bar C}, {\bar E})$,
respectively. For an instance $D$ and a pair of tuples $t_1\in R(D)$ and $t_2\in S(D)$,
$t_1[\bar A]\approx t_2[\bar B]$ indicates that the similarities of the values for all corresponding pairs
of attributes of $(\bar A,\bar B)$ hold. Similarly,  $t_1[\bar C] = t_2[\bar E]$ denotes the
equality of the values of all pairs of corresponding attributes of $({\bar C}, {\bar E})$.

Since an MD involves an update operation,  the MD is a condition that is satisfied by a pair of
database instances: an instance $D$ and its updated instance $D'$.

\begin{definition} \label{def:mdsem} \cite{Fan09} Let $D, D'$ be
instances of schema $\mathcal{S}$ with predicates $R$ and $S$,   such that, for each
tuple $t$ in $D$, there is a unique tuple $t'$
in $D'$ with the same identifier as $t$, and viceversa.
The pair $(D,D')$ satisfies the MD $m$ in (\ref{eq:simplMD}), denoted $(D,D')$
$\vD_F m$, iff, for every pair of tuples
$t_R\in R(D)$ and $t_S\in S(D)$, if $t_R$ and
$t_S$ satisfy  $t_R[{\bar A}]\approx t_S[{\bar B}]$, then for the corresponding
tuples $t_R'$ and $t_S'$ in $R(D'), S(D')$, resp., it holds: (a) $t_R'[\bar C] = t_S'[\bar E]$, and
(b) $t_R'[\bar A] \approx t_S'[\bar B]$. \boxtheorem
\end{definition}
Intuitively, $D'$ in Definition \ref{def:mdsem} is an instance
obtained from $D$ by enforcing $m$ on instance $D$. For a set $M$ of MDs,
and a pair of instances $(D,D')$,
$(D,D')\vD_F M$ means that  $(D,D')\vD_F m$, for every $m\in M$.

An instance $D'$ is {\em stable}  \cite{Fan09} for a set $M$ of MDs if $(D',D')$
$\vD_F M$. Stable instances correspond to the intuitive notion of a clean
database, in the sense that all the expected value identifications
already take place in it.
Although not explicitly developed in \cite{Fan09}, for an instance $D$, if $(D,D')\vD_F M$ for a stable
instance $D'$, then $D'$ is expected to be reached as a fix-point of an iteration of value
identification
updates that starts from $D$ and is based on $M$.

\section{MD Semantics Revisited}\label{sec:prob}

Condition (b) in Definition \ref{def:mdsem} is used to avoid
that the identification updates  destroy the original similarities.
Unfortunately,
enforcing the requirement sometimes leads to
counterintuitive results.

\begin{example}
Consider the following instance $D$ with  string-valued attributes, and MDs:

\begin{center}
\begin{tabular}{l|c|c|c|}
$R$ & $A$ & $B$ & $C$\\ \hline
&$a$ & $c$ & $g$ \\
&$a$ & $c$ & $\nit{ksp}$
\end{tabular}
~~~
\begin{tabular}{l|c|c|}
$S$ & $E~$ & $F$\\ \hline
& $h$~ & $c$\\
& \nit{msp} & $c$
\end{tabular}
\end{center}

\begin{eqnarray}
R[A]\approx R[A] &\ra& R[C]\rl R[C]\\
R[C]\approx S[E] &\ra& R[B]\rl S[F] \label{eq:ex}
\end{eqnarray}

\noindent  For two
strings $s_1$ and $s_2$, $s_1\approx s_2$ if
the edit distance $d$ between $s_1$ and $s_2$
satisfies $d\leq 1$. To produce an instance
$D'$ satisfying $(D,D')\vD_F M$, the
strings $g$ and $\nit{ksp}$ must be changed to some
common string $s'$.

 Because of the similarities
$h\approx g$ and $\nit{ksp} \approx \nit{msp}$, \ $s'$ must be
similar to the $E$ attribute values of the
tuples in $S$, by condition (b) of Definition \ref{def:mdsem} and
MD (\ref{eq:ex}).
Clearly, there is no $s'$ that is similar
to both $h$ and $\nit{msp}$. Therefore, at least one of $h$ and
$\nit{msp}$ \ must be modified to some new value in $D'$.
\boxtheorem
\end{example}
Another problem with the semantics of MDs
is that it allows duplicate resolution in
instances that are already resolved. Intuitively,
there is no reason to change the values in an
instance that is stable for a set of
MDs $M$, because there is no reason to believe,
on the basis of $M$, that these values are in
error. However, even if an instance $D$ satisfies $(D,D)\vD_F M$,
 it is always possible,
by choosing different common values, to
produce a different instance $D'$ such that
$(D,D')\vD_F M$. This is illustrated in the
next example.

\begin{example}
Let $D$ be the instance below and  the MD $R[A]\approx R[A]\ra R[B]\rl R[B]
$.

\begin{center}
\begin{tabular}{l|c|c|}
$R$&$A$ & $B$   \\ \hline
&$a$ & $c$   \\
&$a$ & $c$
\end{tabular}
\end{center}
\noindent Although $D$ is stable, $(D,D')\vD_F m$ is
true for any $D'$ where the $B$ attribute
values of the two tuples are the same. \boxtheorem
\end{example}

\subsection{MD satisfaction}

We now propose a new semantics for MD satisfaction that
disallows unjustified attribute modifications. We keep condition (a)
of Definition \ref{def:mdsem}, while replacing
condition (b) with a restriction on the
possible updates that can be made.

\begin{definition}\label{def:mod}
Let $D$ be an instance of
schema $\mathcal{S}$,  $R \in \mathcal{S}$,
$t_R \in R(D)$, $C$ an attribute of $R$, and $M$ a set of MDs.
Value $t_R[C]$ is \textit{modifiable}
if there exist $S\in \mathcal{S}$,
$t_S\in S(D)$, an $m\in M$ of the form
$R[\bar A]\approx S[\bar B]\ra R[\bar C]\rl S[\bar E]$,
and a corresponding pair $(C,E)$
of $(\bar C,\bar E)$, such that
one of the following holds: 1. $t_R[\bar A]\approx t_S[\bar B]$, but
$t_R[C]\neq t_S[E]$. \ \ 2. $t_R[\bar A]\approx t_S[\bar B]$ and $t_S[E]$ is modifiable.
\boxtheorem
\end{definition}

\begin{example}\label{ex:mod}
Consider two relations $R$ and $S$ with two MDs defined on them:
\vspace{-1mm}
\begin{center}
\begin{tabular}{l|c|c|}
$R$ & $A$ & $B$ \\ \hline
$t_0$ & $a_0$ & $b$ \\
$t_1$ & $a_1$ & $b$ \\
$t_2$ & $a_2$ & $b$
\end{tabular}
~~~
\begin{tabular}{l|c|c|}
 $S$ & $C~$ & $E$\\ \hline
 $t_3$ & $a_3$ & $c$\\
 $t_4$ & $a_4$ & $c$\\
 $t_5$ & $a_5$ & $c$
\end{tabular}
\end{center}

\begin{eqnarray}
m_1:~R[A]\approx R[A] &\ra& R[B]\rl R[B],\nn\\
m_2:~R[A]\approx S[C] &\ra& R[B]\rl S[E].\nn \label{eq:ex2}
\end{eqnarray}

\noindent The following similarities
hold on the distinct constants of $R$ and $S$:
$a_i\approx a_{(i+1)mod6}$, $0\leq i\leq 5$. The
values $t_2[B]$ and $t_3[E]$ are modifiable by
condition 1 of Definition \ref{def:mod}, $m_2$, $a_2\approx a_3$, and
$t_2[B]\neq t_3[E]$. For the same reason, $t_0[B]$
and $t_5[E]$ are modifiable.

 Value $t_1[B]$ is
modifiable by condition 2 of Definition \ref{def:mod},
$m_1$, $a_1\approx a_2$, and the fact that $t_2[B]$
is modifiable. Similarly, $t_4[E]$ is modifiable. \boxtheorem
\end{example}

\begin{definition}\label{def:new}
Let $D$, $D'$ be instances for $\mathcal{S}$ with the same tuple ids, $M$ a set of MDs, and $m \in M$.
$(D,D')$ satisfies $m$, denoted $(D,D')\vD m$, iff:\\
1. For any pair of tuples $t_R\in R(D)$,
$t_S\in S(D)$, if there exists an MD in $M$ of the form
$R[\bar A]\approx S[\bar B]\ra R[\bar C]\rl S[\bar E]$ and
$t_R[\bar A]\approx t_S[\bar B]$, then for the
corresponding tuples $t_R'\in R(D')$ and $t_S'\in S(D')$,
it holds $t_R'[\bar C] = t_S'[\bar E]$.\\
2. For any tuple $t_R\in R(D)$
and any attribute $G$ of $R$,
if $t_R[G]$ is not modifiable, then
$t_R'[G] = t_R[G]$. \boxtheorem
\end{definition}
Notice that the notion of satisfaction of an MD is relative to a set of MDs to which the former belongs (due to the
modifiability condition). Of course, for a single MD $m$, we can consider the set $M = \{m\}$. Condition
2. captures a natural default condition of persistence of values: those that have to be changed are changed only.

The definition
of satisfaction of a set $M$ of MDs, $(D,D)' \models M$,  is as usual. Also, as before, we define
{\em stable} instance for $M$ to mean
$(D,D)\vD M$. Except where otherwise noted, these are the notions of satisfaction and stability that we will use in the
rest of this paper.

\begin{example}
Consider again example \ref{ex:mod}. The set
of all $D'$ such that $(D,D')\vD M$ is the set of all instances
obtained from $D$ by changing all values of
$R[B]$ and $S[E]$ to a common value, and leaving
all other values unchanged. This is because
the values of $R[B]$ and $S[E]$ are the only
modifiable values, and these values must be
equal by condition 1 of Definition \ref{def:new}
and the given similarities. \boxtheorem
\end{example}
Condition 2 in Definition \ref{def:new} on the set of updatable values does not prevent
us from obtaining instances $D'$ that enforce the MD, as the following theorem establishes.

\begin{theorem}\label{thm:arb}
For any instance $D$ and set of MDs $M$,
there exists a $D'$ such that $(D,D')\vD M$.
Moreover, for any attribute value that is
changed from $D$ to $D'$, the new
value can be chosen arbitrarily, as long as
it is consistent with $(D,D')\vD M$. \boxtheorem
\end{theorem}
The new semantics introduced in Definition \ref{def:new}
solves the problems mentioned at the beginning of this section.
Notice that it does not
require additional changes to preserve similarities (if the original ones were
broken). Furthermore, modifications of instances, unless required by the
enforcement of matchings as specified by the MDs, are not allowed.
Also notice that the instance $D'$ in Theorem \ref{thm:arb} is not guaranteed to
be stable. We address this issue in the next section.

Moreover, as can be seen from the proof of
Theorem \ref{thm:arb}, the new restriction
imposed by Definition \ref{def:new} is as strong
as possible in the following sense: Any definition of
MD satisfaction
that includes condition 1. must allow the modification
of the modifiable attributes (according to Definition \ref{def:mod}).
Otherwise, it is not possible to ensure, for arbitrary $D$, the existence of an instance $D'$ with
$(D,D')\vD M$.

\subsection{Resolved instances}

According to the MD semantics in \cite{Fan09}, although not explicitly stated there, a  clean version $D'$ of an instance $D$ is  an instance $D'$ satisfying the  conditions
$(D,D')\models M$ and $(D',D')\models M$. Due to the natural restrictions
on updates captured by the new semantics (cf. Definition \ref{def:new}), the existence of
such a $D'$ is not guaranteed. Essentially, this is because $D'$ is
the result of a series of updates. The MDs are applied to the original instance $D$ to produce a
new instance, which may have new pairs of similar values, forcing another
application of the MDs, which in their turn produces another instance, and so on, until
a stable instance $D'$ is reached. The pair $(D,D')$ may not satisfy $M$.
However, we will be interested in those instances $D'$ just mentioned. The idea
is to relax the condition  $(D,D')\vD M$,
and obtain a stable $D'$ after
 an iterative process of MD enforcement, which at each step, say $k$, makes sure that $(D_{k-1},D_k)
 \models M$.

\begin{definition}\label{def:res}
Let $D$ be a database instance and
$M$ a set of MDs. A {\em resolved instance} for
$D$ wrt $M$ is an instance $D'$,
such that there is a finite (possibly empty) sequence of instances
$D_1,D_2,...D_n$ with:
$(D,D_1)\vD M$, $(D_1,D_2)\vD M$,...
$(D_{n-1},D_n)\vD M$, $(D_n,D')\vD M$, and
$(D',D')\vD M$.
\boxtheorem
\end{definition}

Note that, by Definition \ref{def:new},
for an instance $D$ satisfying $(D,D)\models M$, it holds $(D,D')\models M$ if
and only if $D' = D$. In this case,
 the only possible set of intermediate instances
is the empty set and $D$ is the only
resolved instance. Thus,
a resolved instance cannot be obtained by making changes
to an instance that is already resolved.

\begin{theorem} \label{theo:resolv}
Given an instance $D$ and a
set $M$ of MDs, there always exists a resolved
instance of $D$ with respect to $M$.
\boxtheorem \end{theorem}

\begin{example}\label{ex:pair}
Consider the following instance $D$ of a relation $R$ and
set $M$ of MDs:

\begin{center}
\begin{tabular}{l|c|c|c|}
$R(D)$ & $A$ & $B$ & $C$ \\ \hline
&$a$ & $b$ & $d$ \\
&$a$ & $c$ & $e$ \\
&$a$ & $b$ & $e$
\end{tabular}
\end{center}
\begin{eqnarray}
R[A]\approx R[A] &\ra& R[B]\rl R[B],\nn\\
R[B]\approx R[B] &\ra& R[C]\rl R[C]. \nn\label{eq:ex3}
\end{eqnarray}
All pairs of distinct constants in $R$ are dissimilar. Two
resolved instances $D_1$ and $D_2$ of $R$ are shown.

\begin{center}
\begin{tabular}{l|c|c|c|}
$R(D_1)$ & $A$ & $B$ & $C$ \\ \hline
&$a$ & $b$ & $d$ \\
&$a$ & $b$ & $d$ \\
&$a$ & $b$ & $d$
\end{tabular}
~~~
\begin{tabular}{l|c|c|c|}
$R(D_2)$ & $A$ & $B$ & $C$\\ \hline
& $a$ & $b$ & $e$\\
& $a$ & $b$ & $e$\\
& $a$ & $b$ & $e$
\end{tabular}
\end{center}

\noindent Notice that $(D,D_1)\not \models M$, because the value
of the $C$ attribute of the second tuple is not
modifiable in $D$. \boxtheorem
\end{example}
The notion of resolved instance is one step towards the characterization of the intended
clean instances. However, it still leaves room for refinement. Actually, the resolved instances
that are of most interest for us are those that are somehow closest to
the original instance. This consideration
leads to the concept of {\em minimal resolved
instance}, which uses as a measure of
change the number of
values that were modified to obtain the clean database. In Example \ref{ex:pair}, instance $D_2$ is a minimal
resolved instance, whereas $D_1$ is not.

\begin{definition}
Let $D$ be an instance.\\
(a) $T_D := \{(t,A)~|~ t \mbox{ is the id of a tuple in } D \mbox{ and } A \mbox{ is an}$\\
$\mbox{attribute of the tuple}\}$.\\
(b) $f_D: T_D \rightarrow U$ is given by: $f_D(t,A) :=
\mbox{ the value for } A$\\
$\mbox{ in the tuple in } D \mbox{  with id } t$.\\
(c) For an instance $D'$ with the same tuple ids as $D$:\\
\hspace*{5mm}$S_{D,D'} := \{(t,A)\in T_D~|~ f_D(t,A) \neq f_{D'}(t',A)\}$. \boxtheorem
\end{definition}
Intuitively, $S_{D,D'}$ is the set of all
values changed in going from $D$ to $D'$.

\begin{definition} \label{def:minim}
Let $D$ be an instance and $M$ a set
of MDs. A {\em minimally resolved instance} (MRI) of $D$
wrt $M$ is a resolved instance
$D'$ such that $|S_{D,D'}|$ is minimum, i.e.
there is no resolved
instance $D''$ with $|S_{D,D''}| < |S_{D,D'}|$. We denote by
$\nit{Res}(D,M)$ the set of minimal resolved instances of $D$ wrt
the set $M$ of MDs.
\boxtheorem
\end{definition}

\begin{example}\label{ex:mri}
Consider the instance below and the MD
$R[A]\approx S[C]\ra R[B]\rl S[D]$.

\vspace{-2mm}
\begin{center}
\begin{tabular}{l|c|c|}
$R$& $A$ & $B$    \\ \hline
&$a_1$ & $b_1$   \\
\end{tabular}
~~~~~~~~
\begin{tabular}{l|c|c|}
$S$ &$C$ & $D$  \\ \hline
&$c_1$ & $d_1$ \\
\end{tabular}
\end{center}

\noindent Assuming that $a_1\approx c_1$, this
instance has two minimal resolved instances, namely

\begin{center}
\begin{tabular}{l|c|c|}
$R$&$A$ & $B$    \\ \hline
&$a_1$ & $d_1$   \\
\end{tabular}
~~~~~~~
\begin{tabular}{l|c|c|}
$S$&$C$ & $D$  \\ \hline
&$c_1$ & $d_1$ \\
\end{tabular}
\end{center}

\begin{center}
\begin{tabular}{l|c|c|}
$R$&$A$ & $B$    \\ \hline
&$a_1$ & $b_1$   \\
\end{tabular}
~~~~~~~
\begin{tabular}{l|c|c|}
$S$&$C$ & $D$  \\ \hline
&$c_1$ & $b_1$
\end{tabular}
\end{center}
\vspace{-0.5cm}\boxtheorem
\end{example}
Considering that MDs concentrate on changes of attribute values, we consider that this notion of
minimality is appropriate. The comparisons have to be made at the attribute value level. Notice that in
CQA a few other notions of minimality and comparison of instances have been investigated \cite{B2006}.

\subsection{Resolved answers}\label{sec:ra}

Let $\mathcal{Q}(\bar{x})$ be a query expressed in the first-order language $L(\mathcal{S})$ associated to schema $\mathcal{S}$. Now we are in position to characterize the admissible answers to $\mathcal{Q}$ from $D$, as those that are invariant
under the matching resolution process.

\begin{definition}  A tuple of constants $\bar{a}$ is a {\em resolved answer} to $\mathcal{Q}(\bar{x})$ wrt the set
$M$ of MDs, denoted
$D \models_{M} \mathcal{Q}[\bar{a}]$, iff  $D' \models \mathcal{Q}[\bar{a}]$, for every $D' \in \nit{Res}(D,M)$.
We denote with $\nit{Res\!An}(D,\mathcal{Q},M)$ the set of resolved answers to $\mathcal{Q}$ from $D$ wrt $M$.
\boxtheorem
\end{definition}
\begin{example} (example  \ref{ex:mri} continued)
The set of resolved answers to the query $\mathcal{Q}_1(x,y): R(x,y)$
is empty since there are no
tuples that are in the instance of $R$ in
all minimal resolved instances.
On the other hand, the set of resolved answers
to the query $\mathcal{Q}_2(x): \exists y(R(x,y)\wedge (y = b_1 \vee y = d_1)$
is $\{a_1\}$. \boxtheorem
\end{example}

In Section \ref{subsec:com} we will study the complexity of the problem of
computing the resolved answers, which we now formally introduce.

\begin{definition}
Given a schema $\mathcal{S}$, a query $\mathcal{Q}(\bar{x}) \in L(\mathcal{S})$, and a set $M$
of MDs, the {\em Resolved Answer Problem} (RAP) is the problem of deciding membership of the set
\bea
\nit{RA}_{\mathcal{Q},M} &:=&
\{(D,\bar{a})~|~\bar{a} \hbox{ is a resolved answer to }
\mathcal{Q} \mbox{ from}\nn\\
&&\hspace*{1.4cm}\hb{instance } D \hb{ wrt } M\}.\nn
\eea
If $\mathcal{Q}$ is a
boolean query, it is the problem of
determining whether $\mathcal{Q}$ is true in all
minimal resolved instances of $D$.
\boxtheorem
\end{definition}

\section{Computing Resolved Instances and Answers}\label{subsec:com}

\ignore{
\comj{I removed all mention of MDs whose graphs are
DAGS from this section, since the proof of that theorem
was incorrect. I also removed the dicussion of
bounds on the number of intermediate states (the $k_D$)
since it is no longer relevant. I replaced the $\Pi^P_2$
complexity bound with a co-NP bound that applies only to
HSC sets.} }

In this section, we consider the complexity of
the $\nit{RA}_{\mathcal{Q},M}$ problem introduced in the
previous section. For this goal it is useful to associate a graph to the set of MDs.
We need a few notions before introducing it.
\begin{definition}
A set $M$ of MDs is in {\em standard form}
if no two MDs in $M$ have the same expression
to the left of the arrow.
\boxtheorem
\end{definition}
Notice that any set of MDs can be put in standard form
by replacing subsets of MDs of the form
$\{
R[\bar A]\approx S[\bar B]\ra R[\bar C_1]\rl S[\bar E_1], \ldots,
R[\bar A]\approx S[\bar B]\ra R[\bar C_n]\rl S[\bar E_n]\}$
by the single MD $
R[\bar A]\approx S[\bar B]\ra R[\bar C]\rl S[\bar E]
$,
where the set of corresponding pairs of attributes of $(\bar C, \bar  E)$
is the union of those of $(\bar C_1, \bar  E_1)$, ...$(\bar C_n, \bar  E_n)$.
From now on, we will assume that all sets of MDs are in standard form.

For an MD $m$,
 LHS$(m)$ and RHS$(m)$ denote
the sets of attributes that appear
to the left side and to right side of the arrow,
respectively.
\begin{definition}
Let $M$ be a set of MDs in standard form. The {\em MD-graph} of
$M$, denoted $\nit{MDG}(M)$, is a directed graph with a vertex labeled
$m$ for each $m\in M$, and with an edge from
$m_1$ to $m_2$ iff \ RHS$(m_1)\bc \ $LHS$(m_2)\neq \emptyset$.
\boxtheorem
\end{definition}

\begin{example}
Consider the set of MDs: \ $m_1:~R[A]\approx S[B]\ra R[C]\rl S[D]$.
$m_2:~R[C]\approx S[D]\ra R[A]\rl S[B]$. \ \
$m_3:~S[E]\approx S[B]\ra T[F]\rl T[F]$. It has the MD-graph shown in Figure 1.
\boxtheorem
\end{example}
\begin{figure}
\label{fig}
\centering
\epsfig{file=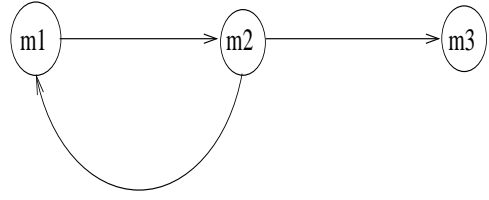, height=1in, width=2.5in}
\caption{An MD-Graph}
\end{figure}
A set of MDs whose MD-graph contains edges is
called {\em interacting}. Otherwise, it is
{\em non-interacting}.

\ignore{
In the following,
we will first establish that, for certain classes  of
MDs whose MD-graphs satisfy some conditions in terms of
cycles, a constant bound can be
derived for the number of steps required to
obtain an MRI.

\vspace{2mm}\noindent {\em Notation:} For a given set $M$ of MDs, an instance $D$, and a minimal resolved instance $D'$,
we denote with $k_{DD'}$ the number of intermediate instances
in Definition \ref{def:res} that are needed to
go from $D$
to $D'$. Also
$k_D$ denotes $\nit{max} \{k_{DD'}|D'$\\
$\hbox{ is an MRI of }D\}$
if this maximum exists.

Clearly, if $(D,D) \models M$, then $k_D = 0$. However, it is not clear
that $k_D$ always exists since there might be no upper bound for the set.
This is because
values outside of
the active domain of $D$ could be introduced into
intermediate states, allowing a possibly
infinite set of MRIs. We now describe sets of MDs
for which $k_D$ exists, and derive bounds for
$k_D$ which are independent of the instance $D$.

\begin{theorem}\label{thm:dag}\em
Let $M$ be a set of MDs and $D$ an
instance. If the MD-graph $\nit{MDG}(M)$ is a
 directed acyclic
graph (DAG), then $k_D\leq p$, where
$p$ is the maximum length of a path in
$\nit{MDG}(M)$.
\boxtheorem \end{theorem}
Later in this section we will use this result to
show, for such a class of MDs and under a reasonable assumption on the
size of values in the database, that $\nit{RA}_{\mathcal{Q},M}$
belongs to $\Pi^P_2$ complexity class \cite{papadimitriou94}.
}

\begin{definition}
(a) A cycle $C$ in an MD-graph $\mdg$ is called
a {\em simple cycle} if
 for each pair $(m_1,m_2)$ of successive
vertices in $C$, the corresponding pairs to the left of the arrow in $m_2$ are corresponding pairs
to the right of the arrow in $m_1$, and do not occur elsewhere in $m_1$.\\
(b) A set $M$ of MDs is {\em simple-cycle} if its MD-graph $\mdg$ is
a simple cycle.
\boxtheorem
\end{definition}

\begin{example}
The following is a simple-cycle set of MDs.
\bea
&&m_1: R[A]\approx S[B]\ra R[C,F]\rl S[E,G],\nn\\
&&m_2: R[C]\approx S[E]\w R[F]\approx S[G] \ra R[A]\rl S[B].\nn
\eea
The MD-graph is a cycle, because attributes in RHS($m_2$) are in
LHS($m_1$), and vice-versa. This cycle is a simple cycle, because the
corresponding pairs $(C,E)$ and $(F,G)$ to the right of the
arrow in $m_1$ are corresponding pairs to the left of the arrow
in $m_2$, and vice-versa.
\boxtheorem
\end{example}
For this class of MDs
it is easy
to characterize the form an MRI takes. This is first
illustrated with an example.

\begin{example}\label{ex:cycle}
Consider the  instance $D$ (with tuple ids) and simple-cycle set of MDs.

\begin{center}
\begin{tabular}{l|c|c|c|}
$R$&$A$ & $B$ & $C$  \\ \hline
1&$a_1$ & $d_1$ & $f$  \\
2&$a_2$ & $e_2$ & $g$  \\
3&$b_1$ & $e_1$ & $h$  \\
4&$b_2$ & $d_2$ & $i$
\end{tabular}
\end{center}
\bea
R[A]\approx R[A]\ra R[B]\rl R[B],\nn\\
R[B]\approx R[B]\ra R[A]\rl R[A].\nn
\eea
The only similarities are: $a_i \approx a_j, \
b_i \approx b_j, \ d_i \approx d_j, \ e_i \approx e_j$, with $i,j \in
\{1,2\}$. If the MDs are applied twice,
successively, to the instance, one
possible result is:
\begin{center}
\begin{tabular}{l|c|c|c|}
&$A$ & $B$ & $C$   \\ \hline
1&$a_1$ & $d_1$ & $f$ \\
2&$a_2$ & $e_2$ & $g$  \\
3&$b_1$ & $e_1$ & $h$  \\
4&$b_2$ & $d_2$ & $i$
\end{tabular}
~~~$\ra$~~~
\begin{tabular}{l|c|c|c|}
&$A$ & $B$ & $C$ \\ \hline
1&$b_2$ & $d_1$ & $f$\\
2&$a_2$ & $d_1$ & $g$\\
3&$a_2$ & $e_1$ & $h$  \\
4&$b_2$ & $e_1$ & $i$
\end{tabular}
\end{center}

\begin{center}
~~~~~~~~~~~~~~~~~~~~~~~~~~~~$\ra$~~~
\begin{tabular}{l|c|c|c|}
&$A$ & $B$ & $C$ \\ \hline
1&$a_2$ & $e_1$ & $f$\\
2&$a_2$ & $d_1$ & $g$\\
3&$b_2$ & $d_1$ & $h$ \\
4&$b_2$ & $e_1$ & $i$
\end{tabular}
\end{center}
\noindent From this it is clear that,
in any sequence of states
$D,D_1,D_2,$ $...$ obtained by
applying the MDs, the updated
instances must have the following pairs
of values equal:
{\small \begin{center}
\begin{tabular}{|c|c|c|}\hline
$D_i$, $i$ odd & \multicolumn{2}{c}{Column}\vline \\ \hline
 & $A$ & $B$    \\ \hline
tuple (id) pairs & $(1,4)$, $(2,3)$ & $(1,2)$, $(3,4)$ \\ \hline
\end{tabular}
~~~~~
\begin{tabular}{|c|c|c|}\hline
$D_i$, $i$ even & \multicolumn{2}{c}{Column}\vline \\ \hline
 & $A$ & $B$    \\ \hline
tuple (id) pairs & $(1,2)$, $(3,4)$ & $(1,4)$, $(2,3)$ \\ \hline
\end{tabular}
\end{center}}
\noindent In any stable instance, the pairs of values in the
above tables must be equal. Clearly, this can only be
the case if all values in the $A$ and $B$ columns
are equal. This can be achieved with a single
update, choosing any value as the common value.
Thus, the MRIs of any instance are those with all values
in the $A$ and $B$ columns set to their most common
value. In the case of  $D$ above, there are 16
MRIs.
\boxtheorem
\end{example}
The Algorithm {\em ComputeMRI} below generalizes the idea presented
in Example \ref{ex:cycle}. It computes the set of all
MRIs for the case of an arbitrary simple
cycle. (The relevant definitions are given
below).

\begin{definition}\label{def:tc}
Let $m$ be the MD
$R[\bar A]\approx S[\bar B]\ra R[\bar C]\rl S[\bar E]$.
The transitive closure, $T^\approx$, of $\approx$ is
the transitive closure of the binary relation relation on tuples
$t_1[\bar A]\approx t_2[\bar B]$, where $t_1\in R$ and
$t_2\in S$.
\boxtheorem
\end{definition}
\ignore{
\comlb{Why only one MD? Should'n it be them $T_m$? What if more MDs?
What is $t$ above? To which relation does it belong? A short example would help.}
\comj{The case of more than one MD is covered by definition \ref{def:tcset}. I
used two definitions because I thought it would be simpler. I changed the
definition in response to this comment.}  }
Notice that Definition \ref{def:tc} implies that
the transitive closure of $\approx$ is an equivalence
relation on the tuples of $R$ and $S$. It therefore
forms a partition of these tuples into disjoint
equivalence classes.
\begin{definition}\label{def:tcset}
For a set $S$ of
binary relations, the transitive
closure, $T^S$, of $S$ is the transitive closure of
the union of all relations in $S$.
\boxtheorem
\end{definition}
This definition can be applied, in particular, to the $T^\approx$s in
Definition \ref{def:tc}, for several MDs.  For the case in which $S$ in Definition \ref{def:tcset}
is a set of equivalence relations, $T^S$ is also an
equivalence relation. These definitions are used in Algorithm {\em ComputeMRI}
in Table \ref{tab:alg}.

\ignore{
\comlb{You shouldn't use the same $T$ as in the previous def. Check. Are all the
binary relations over the same domain? If yes, can we say just the TC of the union
of them? Do you need reflexivity, symmetry too? Clarify all this.}
\comj{I changed the definition in response to this comment. Reflexivity and
symmetry are satisfied when the relations in $S$ have these properties,
which is always the case in this paper.}  }

\begin{table}
\centering
\caption{Algorithm {\em ComputeMRI}}
\begin{tabular}{|p{8cm}|}\hline
\label{tab:alg}
\textbf{Input}: A database instance $D$ and a simple-cycle set $M$ of MDs.\\
\textbf{Output}: Set of MRIs of $D$ with respect to $M$.\\
1)~\quad \textbf{For} \rm $1\leq j\leq n$\\
2)~\quad\qquad Compute $T_j$, the transitive closure of $\approx_j$\\
3)~\quad Compute the transitive closure $T$ of the set $\{T_j|1\leq$\\
~~~\quad $j\leq n\}$\\
4)~\quad\textbf{For} \rm each corresponding pair of attributes $(A,B)$\\
~~~\quad that appears in $M$:\\
5)~\quad\qquad\textbf{For} \rm each equivalence class $E$ defined by\\
~~~~\quad\qquad $T$:\\
6)~\quad\qquad\qquad Choose a value $v$ from among the $A$ and $B$\\
~~~~\quad\qquad\qquad attribute values of tuples in $R\bc E$ and\\
~~~~\quad\qquad\qquad $S\bc E$, respectively, such that no other\\
~~~~\quad\qquad\qquad value occurs more frequently\\
7)~\quad\qquad\qquad\textbf{For} \rm each tuple $t\in E$: \\
8)~\quad\qquad\qquad\qquad $t[A]\lar v$ if $t\in R$ \\
9)~\quad\qquad\qquad\qquad $t[B]\lar v$ if $t\in S$ \\
10)~\quad Repeat 4-9 for other choices of $v$ to produce other\\
~~~~~\quad MRIs\\
11)~\quad Return the resulting set of MRIs\\ \hline
\end{tabular}
\end{table}

\begin{proposition}\label{lem:MRI}
Algorithm {\em ComputeMRI} returns the set of all MRIs
of $D$ wrt a simple-cycle set $M$ of MDs. \boxtheorem
\end{proposition}
\ignore{The following corollary can be obtained from the bound on the
number of updates in the proof of Proposition \ref{lem:MRI}.
\begin{corollary}\label{cor:MRI}\em
For any instance $D$ on which is defined a
simple-cycle set of MDs, $k_D$ is a constant
that depends linearly on the number of MDs.
\end{corollary}}
With some minor modifications to
the $T$ relation in Algorithm {\em ComputeMRI}, we can make the latter work also
for sets
of MDs whose vertices in
the MD-graph can occur on more than one simple
cycle, as shown in Figure 2.\footnote{The modification
involves using the tuple-attribute closure
introduced in Definition \ref{def:taclosure}, for the
cases where the MD-graph has more than one
connected component.}
The following HSC class of sets of MDs extends the simple-cycle class.

\begin{definition} A  set
$M$ of MDs is {\em hit simple cyclic} (HSC) iff each vertex
in $\nit{MDG}(M)$ is on at least one simple cycle of $\nit{MDG}(M)$.
\boxtheorem
\end{definition}
\ignore{It is not hard to show that Corollary \ref{cor:MRI}
applies to HSC sets of MDs. In fact, the following
result can be easily obtained from Theorem \ref{thm:dag} and Proposition
\ref{lem:MRI}.\\
\begin{theorem}\label{cor:DAGHSC}\em
Let $M$ be a set of MDs such that, for  each subset $M'$ corresponding to a connected component of
$\mdg$ it holds:  $\nit{MDG}(M')$ is either a DAG or $M'$ is an HSC set. Then for
any instance $D$, $k_D$ is bounded by a
constant that is linear in the number of vertices of
$\mdg$.
\end{theorem}
}
\begin{figure}
\label{fig2}
\centering
\epsfig{file=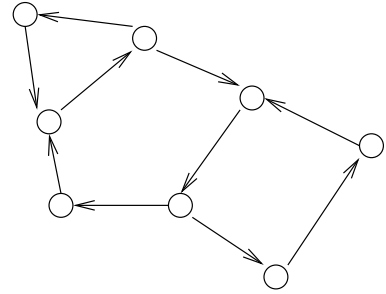}
\caption{The MD-graph of an HSC  set of MDs}
\end{figure}
The next example shows that even for simple classes of MDs, there may be exponentially many MRIs.

\begin{example} \label{ex:exp}
Consider the relational predicate $R[A,B]$ and
the MD $m: R[A]\approx R[A]\ra R[B]\rl R[B]$.
Let $D$ be an instance of $R$ with tuples $\{t_i|~1\leq i\leq n\}$, for some
even number $n$, such that: (a) The values in $D$ satisfy
the similarities $t_i[A]\approx t_{i+1}[A]$ for all odd $i$ with $1\leq i\leq n - 1$, and no others, and
(b) $t_i[B]\neq t_{i+1}[B]$ for all odd $i$ with $1\leq i\leq n - 1$.
 It is clear
that an MRI is obtained by setting the $B$ attributes of $t_i$
and $t_{i+1}$ to either $t_i[B]$ or $t_{i+1}[B]$ for each odd $i$
such that $1\leq i\leq n - 1$. The number of MRIs is the number of
possible choices of such values, which is $2^{n/2}$.
\boxtheorem
\end{example}
The MD in the previous example is HSC.
 Actually, the simple form of the MRIs for HSC sets can be used to obtain an upper bound
for $\nit{RA}_{\mathcal{Q},M}$ that (under usual complexity-theoretic
assumptions) is lower than exponential. This
 relies on the assumption that, if a resolved instance contains
values outside the active domain of the
original instance, then those values are bounded above
in length by a
polynomial in the size of
the original instance. This assumption is in
accord with practical constraints on databases and
any reasonable definition of similarity.

\begin{theorem}\label{theo:hsc}
For HSC sets of MDs, if
resolved instances are restricted to
contain values bounded in length by a
polynomial in the length of the input, then
problem $\nit{RA}_{\mathcal{Q},M}$ is in $\nit{coNP}$ for
any first-order query $\mathcal{Q}$.
\boxtheorem \end{theorem}
In this section, we established a complexity bound for $\nit{RA}_{\mathcal{Q},M}$
which holds
for class of MDs with cyclic MD-graphs and
all first-order queries. The bound follows from the simple form of the MRIs, as
described by Algorithm {\em ComputeMRI}.
In the next section, we further exploit this latter
result to show that, for HSC sets and certain first-order queries, the
resolved answers can be retrieved in polynomial time.

\section{Resolved Query Answering:\\ Tractability and Rewriting}\label{sec:Tr}

In this section, we discuss  tractable cases of $RA_{\mathcal{Q},M}$. In particular, we propose a
query rewriting technique for obtaining the resolved
answers for certain FO queries and MDs. In Section \ref{subsec:NI2}, we will relate
$RA_{\mathcal{Q},M}$ to {\em consistent query
answering} (CQA) \cite{B2006}. This connection and some known results in CQA  will allow us to identify further tractable cases,
but also to establish the
intractability of $RA_{\mathcal{Q},M}$ for certain classes of queries and MDs. The latter makes the tractability
results obtained in this section even more
relevant.

A possible approach to obtaining the resolved
answers to a query $\mathcal{Q}$ from an instance $D$ is to rewrite $\mathcal{Q}$ into a new query $\mathcal{Q}'$ on
the basis of $\mathcal{Q}$ and $M$. $\mc{Q}'$ should be such that, when posed to $D$ (as usual), it
returns the resolved answers to $\mathcal{Q}$ from $D$. In this case, it is not necessary to explicitly
compute the MRIs. If $\mc{Q}'$ can be efficiently evaluated against $D$, then the resolved answers can also
be efficiently computed and $RA_{\mathcal{Q},M}$ becomes tractable. This methodology was proposed in \cite{Arenas99}
for CQA.

This section investigates this query rewriting approach to the computation of resolved answers for
HSC sets of MDs. The input queries $\mc{Q}$
will be conjunctive queries with
certain restrictions on the joins. However,
the rewritten queries $\mc{Q}'$ may involve aggregate operators (actually, $\nit{Count}$), universal quantification, and Datalog
rules (to specify the transitive closure).
We will need to compute
transitive closures and count the number
of occurrences of values in order to enforce
minimal change. In any case, the resulting query $\mc{Q}'$ will still be evaluable
in polynomial time in the size of $D$.

Specifically, the input queries we consider
have the form \ $\mathcal{Q}(\bar x) : \ex\bar u (R_1(\bar v_1) \wedge \cdots \wedge R_n(\bar v_n))$,
 where $\bar{x} = (\cup \bar{v_i}) \smallsetminus \bar{u}$. For tractability of $RA_{\mathcal{Q},M}$, we
 need additional restrictions on them.

\begin{definition}
(a) For a set $M$ of MDs defined on
schema $\mc{S}$,
the {\em changeable
attributes} of $\mc{S}$ are those that appear to the right of the arrow
in some $m \in M$. The other attributes of $\mc{S}$
are called {\em unchangeable}.\\
(b) Let $\mathcal{Q}$ be a conjunctive query
and $M$ a set of MDs. Query $\mathcal{Q}$
is an {\em unchangeable attribute join} conjunctive query \
($\nit{ucajCQ}$) if there are no bound, repeated
variables in $\mathcal{Q}$ that correspond to
changeable attributes. \boxtheorem
\end{definition}

\begin{example} Let $M$ be the single MD
$R[A]\approx R[A] \ra R[B]\rl R[B]
$. The query
$\mathcal{Q}(x,z) : \ex y (R(x,y)\w R(z,y))$
is not in the $\nit{ucajCQ}$ class, because it
contains a bound, repeated variable
($y$) which corresponds to a changeable
attribute ($B$). However, the
query
$\mathcal{Q}(y) : \ex x \ex z (R(x,y)\w R(x,z))$
is in $\nit{ucajCQ}$, since the only bound,
repeated variable ($x$) corresponds to an
unchangeable attribute ($A$). \boxtheorem
\end{example}
In Section \ref{subsec:NI2} we will encounter HSC MDs (even non-interacting MDs) and
conjunctive queries outside $\nit{ucajCQ}$ for which $RA_{\mathcal{Q},M}$ is intractable (cf. Theorem
 \ref{thm:Chom} below).

To incorporate counting into FO queries, we will use the operator $\nit{Count}(R)$ that
returns the number of tuples in relation $R$ (cf. \cite{Abiteboul}).
$\nit{Count}$ will be applied to sets of tuples of the form $\{\bar t~|~C\}$, where
$\bar t$ is a tuple of variables, and
and $C$ is a FO condition
whose free variables include those in
$\bar t$. Now we show a simple example of rewriting that uses $\nit{Count}$.

\begin{example}  \label{ex:count} Consider a
relation $R$, the MD $R[A]\approx R[A]\ra R[B]\rl R[B]$, and the query $\mathcal{Q}(x,y,z) : R(x,y,z)$.
$R$ and its (single) MRI are shown below.

\begin{center}
\begin{tabular}{l|c|c|c|}
$R$&$A$ & $B$ & $C$  \\ \hline
&$a_1$ & $b_1$ & $c_1$  \\
&$a_1$ & $b_2$ & $c_2$  \\
&$a_1$ & $b_2$ & $c_3$
\end{tabular}
~~~~~~
\begin{tabular}{l|c|c|c|}
$\mbox{MRI}$&$A$ & $B$ & $C$  \\ \hline
&$a_1$ & $b_2$ & $c_1$  \\
&$a_1$ & $b_2$ & $c_2$  \\
&$a_1$ & $b_2$ & $c_3$
\end{tabular}
\end{center}

\noindent The set of resolved
answers to $\mathcal{Q}$ is
$\{(a_1,b_2,c_1),(a_1,b_2,c_2),$\\
$(a_1,b_2,c_3)\}$. It is not difficult to see that the following query
returns the resolved answers (for any initial instance of $R$). In it,  $T$ stands for the transitive closure $T^\approx$ of
$\approx$ (cf. Definition \ref{def:tc}).
\bea\label{eq:rewrite}
&& \mathcal{Q}'(x,y,z) : \ex y'R(x,y',z)\wedge\fa y''[\nit{Count}\{
(x',y,z')~|\nn\\
&& T((x,y',z),(x',y,z')) \wedge R(x',y,z')\} >\nit{Count}\{(x',y'',z')~|\nn\\
&& T((x,y',z),(x',y'',z'))\wedge R(x',y'',z')\wedge y''\neq y\}].\nn
\eea
Intuitively, the first conjunct requires the
existence of a tuple $t$ with the same $A$ and
$C$ attribute values as the answer. Since
the values of these attributes are not
changed when going from the original instance
to an MRI, such a tuple must exist. However,
the tuple is not required to have the same
$B$ attribute value as the answer tuple,
because this attribute can be modified.
For example, $(a_1,b_2,c_1)$ is a resolved
answer, but is not in $R$. What makes it a
resolved answer is the fact that it is in
an equivalence class of $T$ (consisting of
all three tuples in $R$) for which $b_2$
occurs more frequently as a $B$ attribute value
than any other value. This condition
on resolved answers is
expressed by the second conjunct.
\boxtheorem
\end{example}
For simplicity, we present our query rewriting
algorithm for non-interacting
MDs, a special case of
HSC sets of MDs where the
connected components have only one
vertex. The generalization to arbitrary
HSC sets is straightforward, and the
required modifications are indicated
at the end of this section. First we
require the following definitions.
\begin{definition}\label{def:atclosure}
Let $M$ be a set of MDs on schema $\mc{S}$.
(a) Define a (symmetric) binary relation $\rl_r$ which relates attributes $R[A]$, $S[B]$ of $\mc{S}$
 if there is an MD in $M$ where $R[A]\rl S[B]$ appears
to the right of the arrow.\\
(b)
The {\em attribute closure}, $T_\nit{at}$, of $M$ is the
binary relation on attributes
defined as the reflexive, transitive closure of
$\rl_r$. \\
(c) We use the notation $E_{R[A]}$ to
denote the equivalence class of $T_\nit{at}$ to
which attribute $R[A]$ belongs.
\boxtheorem
\end{definition}
Note that,
in general, there will be pairs of attributes
$R[A],$ $S[B]$ for which $E_{R[A]} = E_{S[B]}$.

\begin{example}\label{ex:atclosure} Let $M$ be the
set of MDs
\bea
&&R[A]\approx_1 S[B]\ra R[C]\rl S[D],\nn\\
&&S[E]\approx_2 T[F] \w S[G]\approx T[H]\ra
S[D,K]\rl T[J,L],\nn\\
&&T[F]\approx_3 T[H]\ra T[L,N]\rl T[M,P].\nn
\eea
The equivalence classes of $T_{at}$
are $E_{R[C]} = \{R[C],S[D],T[J]\}$,
$E_{S[K]} = \{S[K],$ $T[L],T[M]\}$, and
$E_{T[N]} = \{T[N],T[P]\}$.
\boxtheorem
\end{example}
To describe the MRIs in this case, we need
the transitive closure relation defined below.

\begin{definition}\label{def:taclosure} Let $m$ be the MD
$R[\bar A]\approx S[\bar B]\ra R[\bar C]\rl S[\bar E]$. \\
(a) Let $\approx'$ be the following binary relation on
tuple-attribute pairs: $(t_1,C)\approx'(t_2,E)
\ :\Leftrightarrow \ t_1[\bar A] \ \approx \ t_2[\bar B]$
and $(C,E)$ is a corresponding pair of $(\bar C,\bar E)$.\\
(b) The {\em tuple-attribute closure} $\nit{TA}$ of $m$ is
the reflexive, transitive closure of $\approx'$.
\boxtheorem
\end{definition}
We denote by $\nit{TS}$ the transitive closure of a set of
tuple-attribute closures (cf. Definition \ref{def:tcset}).
$\nit{TS}$ partitions the set of tuple/attribute pairs
into disjoint equivalence classes.

To keep the notation simple, we
omit parentheses delimiting tuples and tuple/attribute pairs when
writing the arguments of $\nit{TA}$ and $\nit{TS}$. For example, for
tuples $t_2 = (a,b,c)$ and $t_3 = (d,e,f)$ with attributes $A$
and $C$, respectively, $\nit{TS}(((a,b,c),A),((d,e,f),C))$ is written
as $\nit{TS}(a,b,c,A,d,e,$\\ $f,C)$.

Algorithm {\em Rewrite} in Table \ref{tab:rewrite},
outputs a rewritten query $\mc{Q}'$ that returns the resolved answers to
a given input conjunctive query $\mc{Q}$ and set of non-interacting MDs.
This is done by separately rewriting each conjunct $R_i(\bar v_i)$
in $\mc{Q}$.
If $R_i(\bar v_i)$ contains no free variables, then
it is unchanged (line 5). Otherwise, it is replaced with
a conjunction involving the same atom and additional
conjuncts which use the $\nit{Count}$ operator. The conjuncts
involving \nit{Count} express the condition that, for each
changeable attribute value returned by the query, this value
is more numerous than any other value in the same set of values
that is equated by the MDs. The \nit{Count} expressions contain new
local variables as well as a new universally quantified variable $v_{iA}''$.

\begin{example}\label{ex:alg}
We illustrate the algorithm with predicates $R[ABC], S[EFG], U[HI]$, the query \
$\mathcal{Q}(x,y,z): \ex t,u,p,q$\\ $(R(x,y,z)\w S(t,u,z)\w U(p,q))$; and the MDs:
$R[A]\approx S[E]\ra R[B]\rl S[F]$ and  $S[E]\approx U[H]\ra S[F]\rl U[I]$.

Since the $S$ and $U$ atoms have no free variables
holding the values of changeable attributes, these
conjuncts remain unchanged (line 5).
The only free variable holding the value of a
changeable attribute is $y$. Therefore, line 7 sets
$\bar v_1'$ to $(x,y',z)$. Variable $y$ contains the value
of attribute $R[B]$.
The equivalence class $E_{R[B]}$ of $T_{at}$
 to which $R[B]$ belongs is
$\{R[B],S[F],U[I]\}$, so the loop at line 11
generates the atoms $R(x',y,z')$, $R(x',y'',z')$,
$S(t',y,z')$, $S(t',y'',z')$, $U(p',y)$, $U(p',y'')$.
The rewritten query is obtained by replacing in $\mc{Q}$ the
conjunct $R(x,y,z)$ by \ $\ex y' (R(x,y',z) \ \w \fa y''[$
\bea
&&\nit{Count}\{(x',y,z')|~TS(x,y',z,R[B],x',y,z',R[B])\w\nn\\
&& R(x',y,z')\} \ + \nit{Count}\{(t',y,z')|~TS(x,y',z,R[B],\nn\\
&& t',y,z',S[F])\w S(t',y,z')\}  + \nit{Count}\{(p',y)|~TS(x,y',z,\nn\\
&& R[B],p',y,U[I])\w U(p',y)\}  \ \ > \nn
\eea

\bea &&\nit{Count}\{(x',y'',z')|\nn\\
&& TS(x,y',z,R[B],x',y'',z',R[B])\w R(x',y'',z')\w  \nn\\
&& y''\neq y\} + \nit{Count}\{(t',y'',z')|~TS(x,y',z,R[B],t',y'',\nn\\
&& z',S[F])\w S(t',y'',z')\w y''\neq y\} +\nit{Count}\{(p',y'')|\nn\\
&& TS(x,y',z,R[B],p',y'',U[I])\w U(p',y'')\w y''\neq y\}]. \hspace{2mm} \Box \nn
\eea
\end{example}

\begin{table}
\centering
\caption{Algorithm {\em Rewrite}}
\begin{tabular}{|p{8cm}|}\hline
\label{tab:rewrite}
\textbf{Input}: A query in $\nit{ucajCQ}$ and non-interacting set of MDs
$M$.\\
\textbf{Output}: The rewritten query $\mathcal{Q}'$.\\
1)~\quad Let $\mathcal{Q}(\bar t) : \ex \bar u \w_{1\leq i\leq n}R_i(\bar v_i)$
be the query.\\
2)~\quad \textbf{For} each $R_i(\bar v_i)$\\
3)~\quad\qquad Let $C$ be the set of changeable attributes of $R_i$\\
~~~~\quad\qquad corresponding to a free variable in $\bar v_i$\\
4)~\quad\qquad \textbf{If} $C$ is empty\\
5)~\quad\qquad\qquad $Q_i(\bar v_i)\lar R_i(\bar v_i)$\\
6)~\quad\qquad \textbf{Else} \\
7)~\quad\qquad\qquad Let $\bar v_i'$ be $\bar v_i$ with each
variable $v_{iA}$ in \\
~~~~\quad\qquad\qquad $\bar v_i$ holding the value of an attribute $A\in C$\\
~~~~\quad\qquad\qquad replaced by a new variable $v_{iA}'$\\
8)~\quad\qquad\qquad Let $\bar v_{iC}$ be the vector of variables $v_{iA}$, \\
~~~~\quad\qquad\qquad $A\in C$\\
9)~\quad\qquad\qquad Let $\bar v_{iC}'$ be the vector of variables $v_{iA}'$,\\
~~~~\quad\qquad\qquad $A\in C$\\
10)\quad\qquad\qquad \textbf{For} each variable $v_{iA}$ in $\bar v_{iC}$\\
11)\quad\qquad\qquad\qquad \bf For \rm each attribute $R_j[B_k]\in E_A$\\
12)\quad\qquad\qquad\qquad\qquad Generate atom $R_j(\bar u_{jk})$, where\\
~~~~\quad\qquad\qquad\qquad\qquad all variables in $\bar u_{jk}$ are new \\
~~~~\quad\qquad\qquad\qquad\qquad except the one holding the value\\
~~~~\quad\qquad\qquad\qquad\qquad of $R_j[B_k]$, which is $v_{iA}$\\
13)\quad\qquad\qquad\qquad\qquad Generate atom $R_j(\bar w_{jk})$, where \\
~~~~\quad\qquad\qquad\qquad\qquad all variables in $\bar w_{jk}$ are labelled\\
~~~~\quad\qquad\qquad\qquad\qquad  as in $\bar u_{jk}$ except the one holding\\
~~~~\quad\qquad\qquad\qquad\qquad the value of $R_j[B_k]$, which is $v_{iA}''$\\
14)\quad\qquad\qquad\qquad\qquad $C_{jk}^{A1}\lar
\nit{Count}\{\bar u_{jk} | TS(\bar v_i',R_i[A],$\\
~~~~\quad\qquad\qquad\qquad\qquad $\bar u_{jk},R_j[B_k])\w R_j(\bar u_{jk})\}$\\
15)\quad\qquad\qquad\qquad\qquad $C_{jk}^{A2}\lar
\nit{Count}\{\bar w_{jk} | TS(\bar v_i',$\\
~~~~\quad\qquad\qquad\qquad\qquad $R_i[A],\bar w_{jk},R_j[B_k])\w R_j(\bar w_{jk})$\\
~~~~\quad\qquad\qquad\qquad\qquad $\w v_{iA}''\neq v_{iA}\}$\\
16)\quad\qquad\qquad $Q_i(\bar v_i)\lar \ex \bar v_{iC}'\{R_i(\bar v_i')\w_{A\in C}\fa v_{iA}''
[\S_{j,k} C_{jk}^{A1}$\\
~~~~\quad\qquad\qquad $> \S_{j,k} C_{jk}^{A2}]\}$\\
17)\quad $\mathcal{Q}'(\bar t)\lar \ex \bar u\w_{1\leq i\leq n}Q_i(\bar v_i)$\\
18)\quad \textbf{return} $\mathcal{Q}'$\\\hline
\end{tabular}
\end{table}

\begin{theorem}\label{thm:rewrite}
For a set $M$ of non-interacting MDs and a query
$\mathcal{Q}$ in the class $\nit{ucajCQ}$, the
query $\mc{Q}'$ computed by Algorithm {\em Rewrite}
returns the resolved answers to $\mc{Q}$ when posed to
any instance.
\boxtheorem
\end{theorem}
As expected, the rewriting algorithm that produced the rewritten query does not depend upon the dirty instance
at hand, but only on the MDs and the input query, and runs in polynomial time.

Algorithm {\em Rewrite} can be easily adapted and extended  to handle
HSC sets of MDs. All that is
required is a modification to the
tuple-attribute closure
in Definition \ref{def:taclosure}, as
 follows: For an
HSC set of MDs $M$ and $m\in M$, a pair of tuples
$t_1$ and $t_2$ satisfies $(t_1,C)\approx'(t_2,E)$
iff $t_1[\bar A]\approx t_2[\bar B]$ and
$(C,E)$ appears as a corresponding pair to the
right of the arrow in some MD in the same
connected component of the MD graph as $m$.
Tuple-attribute closure is redefined as the
transitive closure of this new relation.
As with Theorem \ref{thm:rewrite}, the
correctness proof is based on the simple
form of the MRIs, and is proved using the same
technique as in the proof of Proposition \ref{lem:MRI}.

\section{The CQA Connection}\label{subsec:NI2}

MDs can be seen as a new form of integrity constraint (IC).
An instance $D$ violates an MD $m$ if there are
unresolved duplicates, i.e.  tuples
$t_1$ and $t_2$ in $D$ that satisfy the similarity
condition of $m$, but differ on some pair of attributes
that are matched by $m$. The instances that are
consistent with a set of MDs $M$ are resolved instances
of themselves with respect to $M$. Among classical
ICs, the closest analogues of MDs
are functional dependencies (FDs).

Given a database instance $D$ and a set of ICs $\Sigma$, possibly not satisfied by $D$, consistent query answering (CQA) is the problem of characterizing
and computing the answers to queries $\mc{Q}$ that are true in all
the instances $D'$ that are consistent with $\Sigma$ and minimally differ
from $D$ \cite{Arenas99}. The consistent instances $D'$ are
called {\em repairs}.
 Minimal difference can be defined in different ways. Most of the research in
CQA has concentrated on the case where the symmetric difference of instances, as sets of tuples,
is made minimal under set inclusion \cite{Arenas99,B2006,chom07}. However, also the minimization of the  cardinality of this
difference has been investigated \cite{lopat07,kol09}.  Other forms of minimization measure the differences in
attribute values between $D$ and $D'$ \cite{fran01,wijsen05,fles05,berIS08}.
Because of their practical
importance, much work on CQA has been done
for the case where  $\Sigma$ is a set of functional dependencies (FDs), in particular,
key constraints (KCs) \cite{Chomicki05,fux07,Wijsen07,Wijsen09,Wijsen10}.

Actually, for a set of KCs  $\mc{K}$ and repairs based on tuple deletions,  a {\em repair} $D'$ of an instance $D$ can be characterized
 as a maximal subset of $D$ that satisfies $\mc{K}$: $D' \subseteq D, \ D' \models \mc{K}$ and there is no $D''$ with
 $D' \subsetneqq D'' \subseteq D$, with $D'' \models \mc{K}$ \cite{Chomicki05}.

 Now, for a FO query $\mc{Q}(\bar{x})$ and a set of KCs $\mc{K}$, the {\em consistent query answering problem}
 is about deciding
membership of the set
\bea\label{eq:CQA}
\nit{CQA}_{\mathcal{Q},\mc{K}} = \{(D,\bar{a})~|~\bar{a}\hbox{ is an answer to }
\mathcal{Q}\hb{ in all repairs of }\nn\\
D\hb{ with respect to }\mc{K}\}.\nn
\eea
A $\bar{a}$ satisfying the above
is called a {\em consistent answer} to
 $\mathcal{Q}$ from $D$.

Notice that this notion of minimality involved in repairs wrt FDs  is tuple and set-inclusion
oriented, whereas the one related to MRIs (cf. Definition \ref{def:minim}) is attribute and
cardinality oriented. However,  the connection can still be established. In particular, the following result can be obtained from
\cite[Thm. 3.3]{Chomicki05}.

\begin{theorem}\label{thm:Chom}
Consider a  relational predicate $R[A,B,C]$, the MD
\bea\label{eq:intMD}
m\!: \ R[A] = R[A]\ra R[B,C]\rl R[B,C],
\eea
and the query $\mathcal{Q}\!:
\ex x \ex y \ex y' \ex z(R(x,y,c)\w R(z,y',d)\w y = y')$.
$RA_{\mathcal{Q},\{m\}}$ is $\nit{coNP}$-complete. \boxtheorem
\end{theorem}
Notice that the conjunctive query in this result does not belong to the
$\nit{ucajCQ}$ class.

For certain classes of conjunctive queries and
ICs consisting  of a single KC per relation, CQA has been proved to be tractable.
This is the case for the  $\mathcal{C}_{\!\nit{forest}}$ class
of conjunctive queries
 \cite{fux07}. Actually, for this class there is a FO
 rewriting of the original query that returns
the certain answers.
$\mathcal{C}_{\!\nit{forest}}$ excludes repeated relations and allows joins only
between non-key and key attributes.
Similar results were subsequently proved for a larger
class of queries that includes some queries with
repeated relations and joins between non-key
attributes \cite{Wijsen07,Wijsen09,Wijsen10}. The following result allows us to take advantage of tractability results for CQA in our
MD setting.

\begin{proposition}\label{thm:red}
Let $D$ be a database instance with
a single relation $R$. Let $m$
be a MD of the form
$R[\bar A] = R[\bar A]\ra R[\bar B]\rl R[\bar B]$,
where the set of attributes of $R$
is $\bar A\bigcup \bar B$ and $\bar A\bc \bar B = \emptyset$.
Then there is a polynomial time
reduction from $\nit{RA}_{\mathcal{Q},\{m\}}$ to $\nit{CQA}_{\mathcal{Q},\{\kappa\}}$,
where $\kappa$ is the key constraint $\bar A\ra \bar B$. \boxtheorem
\end{proposition}
Proposition \ref{thm:red} can be easily
generalized to several relations with
one such MD defined on each. The reduction takes an instance $D$ for $\nit{RA}_{\mathcal{Q},\{m\}}$ and produces an instance $D'$ for
$\nit{CQA}_{\mathcal{Q},\{\kappa\}}$. The schema of $D'$ is the same for $D$, but the extensions of
the relational predicates in it are changed wrt $D$ via counting. Since definitions for those aggregations can be included (or inserted) in the query $\mc{Q}$, we obtain:

\begin{theorem} \label{theo:comp}Let $\mathcal{S}$
be a database schema with relation predicates
 $R_i$, $1\leq i\leq n$ with a set $\mc{K}$
 of KCs \
$\kappa_i\!: \ R_i[\bar A_i]\ra R_i[\bar B_i]$, $1\leq i\leq n$.
Let $\mathcal{Q}$ be a FO query, and suppose there
exists a polynomial time computable FO query $\mathcal{Q}'$, such that
$\mathcal{Q}'$ returns the consistent answers to $\mathcal{Q}$
from $D$. Then there exists a polynomial time computable
FO query $\mathcal{Q}''$ with aggregation that
returns the resolved answers to $\mathcal{Q}$ from $D$ wrt the
MDs \
$m_i\!: \ R_i[\bar A_i] = R_i[\bar A_i] \ra R_i[\bar B_i]\rl R_i[\bar B_i],~~1\leq i\leq n$.
\boxtheorem
\end{theorem}
The aggregation in $\mathcal{Q}''$ in Theorem \ref{theo:comp}
arises from the transformation of the instance that
is used in the reduction in Proposition \ref{thm:red}. We emphasize  that
$\mathcal{Q}''$ is {\em not} obtained using algorithm {\em Rewrite}
from Section \ref{sec:Tr}, which is not guaranteed to work
for queries outside the class $\nit{ucajCQ}$. Rather,
a first-order transformation of the $R_i$ relations with
$\nit{Count}$ is composed with $\mathcal{Q}'$ to produce
$\mathcal{Q}''$. Similar to Algorithm {\em Rewrite} in Section \ref{sec:Tr}, they are used to
express the most frequently occurring values for the changeable attributes for a
given set of tuples with identical values for the unchangeable attributes.

This theorem can be applied to decide/compute resolved answers through composition in those cases where a FO rewriting for CQA has been
identified.  In consequence, it extends
the tractable cases identified in Section \ref{sec:Tr}. They can be applied
to queries that are not in $\nit{ucajCQ}$.

\begin{example}
The query $\mathcal{Q}:~\ex x \exists y \exists z (R(x,y)\w S(y,z))$
is in the class $\mathcal{C}_{\!\nit{forest}}$
for relational predicates $R[A,B]$ and $S[C,E]$
and FDs $A\ra B$ and $C\ra E$.
By Theorem \ref{theo:comp} and the results in \cite{fux07},
this implies the existence of a polynomial time computable
FO query with counting that returns the resolved answers
to $\mc{Q}$ wrt  MDs $R[A] = R[A]\ra R[B]\rl R[B]$
and $S[C] = S[C]\ra S[E]\rl S[E]$. Notice that
$\mathcal{Q}$ is not in $\nit{ucajCQ}$, since the bound
variable $y$ is associated with the changeable attribute
$R[B]$. \boxtheorem
\end{example}

\section{Conclusions}\label{sec:con}

In this paper we have proposed a revised semantics for matching dependency (MD) satisfaction
wrt the one originally proposed in \cite{Fan09}. The main outcomes from that semantics are the
notions of {\em minimally resolved instance} (MRI) and {\em resolved answers} (RAs) to queries. The
former capture the intended, clean instances obtained after enforcing the MDs on a given instance.
The latter are query answers that persist across all the MRIs, and can be considered as robust
and semantically correct answers.

We  investigated the new semantics, the MRIs and the RAs. We  considered the existence of
MRIs, their number, and the cost of computing them. Depending on syntactic criteria on MDs and  queries,
tractable and intractable cases of resolved query answering were identified. The tractable cases coincide
with those where the original query can be rewritten into a new, polynomial-time evaluable query that returns
the resolved answers when posed to the original instance. It is interesting that the rewritings make use
of counting and recursion (for the transitive closure). The original queries considered in this paper are all conjunctive. Other classes of queries will be considered in future work.

Many of our results apply to  cases for
which the resolved instances can be
obtained after a single (batch) update operation.
The investigation of cases requiring
multiple updates is a subject of ongoing research.
We have obtained several tractability and intractability results. However, understanding the complexity
landscape requires still much more research.

We established interesting connections between resolved query answering wrt MDs and
consistent query answers. There are still many issues to explore in this direction, e.g. the
possible use
of logic programs with stable model semantics to specify the MRIs, so as it has been done with database repairs \cite{ArenasBC03,BarceloBB01,greco03}.

We have proposed some efficient algorithms for resolved query answering. Implementing them and
experimentation are also left for future work. Notice that those algorithms use different forms of transitive closure.
 To avoid unacceptably slow
query processing, it may be necessary to
compute  transitive closures off-line
and store them.
The use of Datalog with aggregate functions should also be investigated in this direction.

In this paper we have not considered cases where the matchings of attribute values,
whenever prescribed by the MDs' conditions, are made according to matching
functions. This element adds an entirely new dimension to the semantics and the problems investigated here.
 It certainly deserves investigation.

\bibliographystyle{abbrv}

\appendix

\section{Auxiliary Results and Proofs} \label{sec:proofs}

\defproof{Theorem \ref{thm:arb}}{Consider an undirected graph $G$ whose
vertices are labelled by pairs $(t,A)$, where $t$ is
a tuple identifier and $A$ is an attribute of $t$.
There is an edge between two vertices $(s,A)$
and $(t,B)$ iff $s$ and $t$ satisfy the similarity
condition of some MD $m\in M$ such that $A$
and $B$ are matched by $m$.

Update $D$ as
follows. Choose a vertex $(t_1,A)$ such that
there is another vertex $(t_2,B)$ connected to $(t_1,A)$
by an edge and $t_1[A]$ and
$t_2[B]$ must be made equal to satisfy the
equalities in condition 1. of Definition \ref{def:new}.
For convenience in this proof, we say that $t_2$ is
unequal to $t_1$ for such a pair of tuples
$t_1$ and $t_2$. Perform a breadth first
search (BFS) on $G$ starting with $(t_1,A)$ as level 0. During the
search, if a tuple is discovered at level $i+1$
that is unequal to an adjacent tuple at level $i$, the
value of the attribute in the former tuple is modified
so that it matches that of the latter tuple.
When the BFS has completed, another
vertex with an adjacent unequal tuple is chosen and
another BFS is performed. This
continues until no such vertices remain. It is clear that
the resulting updated instance $D'$ satisfies condition 1. of
definition \ref{def:new}.

We now show by induction on the levels of the
breadth first searches that for all vertices
$(t,A)$ visited, $t[A]$ is modifiable. This is true in
the base case, by choice of the starting vertex.
Suppose it is true for all levels up to and
including the $i^{th}$ level. By definition
of the graph $G$ and condition 2. of
definition \ref{def:mod}, the statement is
true for all vertices at the $(i+1)^{th}$
level. This proves the first statement of the
theorem.

To prove the second statement, we show that,
to satisfy condition 1.  of Definition \ref{def:new},
the attribute values represented by each vertex in each
connected component of $G$ must be changed to
a common value in the new instance. The statement
then follows from the fact that the update
algorithm can be modified so that the attribute
value for the initial vertex in each BFS
is updated to some arbitrary value at the start
(since it is modifiable). By condition 1.  of
Definition \ref{def:new}, the pairs of values that
must be equal in the updated instance $D'$
correspond to those vertices that are connected
by an edge in $G$. This fact and transitivity of equality
imply that all attribute values in a connected
component must be updated to a common value.}

\vspace{2mm}
\defproof{Theorem \ref{theo:resolv}}{We give an algorithm to
compute a resolved instance, and use a
monotonicity property to show that it
always terminates.
For attribute domain $d$ in $D$, consider the
set $S^d$ of pairs $(t,A)$ such that
attribute $A$ of the tuple with identifier $t$
has domain $d$. Let $\{S_1, S_2, ... S_n\}$
be a partition of $S^d$ into
sets such that all tuple/attribute pairs in a set
have the same value in $D$.
 Define the level of $(t,A)$ to
mean $|S_j|$ where $(t,A)\in S_j$.

The algorithm first applies all MDs in $M$ to $D$
by setting equal pairs of unequal values according
to the MDs. Specifically, consider a
connected component $C$ of the graph in the
proof of Theorem \ref{thm:arb}. If the values of $t[A]$
for all pairs $(t,A)$ in $C$ are not all the same,
then their values are modified to a common value
which is that of the pair with the highest level. This
update is allowed by Theorem \ref{thm:arb}.
In the case of a tie, the common value is chosen
as the largest of the values
according to some total ordering of the
values from the domain that occur in the instance.
It is easily verified that this operation
increases the sum over all the levels of the
elements of $S^d$, where $d$ is the domain of
the attributes of the pairs in $C$. These updates produce an instance
$D_1$ such that $(D,D_1)\vD M$.

The MDs of $M$ are then applied to the
instance $D_1$ to obtain a
new instance $D_2$ such that $(D_1,D_2)\vD M$ and so on,
until a stable instance is reached. For each new instance,
the sum over all domains $d$ of the levels
of the $(t,A)\in S^d$ is greater than for
the previous instance. Since this
quantity is bounded above, the algorithm
terminates with a resolved instance.}

\vspace{2mm}
For the proof of Proposition \ref{lem:MRI}, we need an auxiliary result.

\begin{lemma}\label{lem:tc}
Let $D$ be an instance and let
$m$ be the MD in Definition \ref{def:tc}.
Let $T$ be the transitive closure
of $\approx$. An instance $D'$ obtained by
changing modifiable attribute values of $D$
satisfies $(D,D')\vD m$ iff
for each equivalence
class of $T$, there is a constant vector
$\bar v$ such that,
for all tuples $t$ in the equivalence class,
\bea
t'[\bar C] = \bar v~~\hbox{if } t\in R(D)\nn\\
t'[\bar E] = \bar v~~\hbox{if } t\in S(D)\nn
\eea
where $t'$ is the tuple in $D'$ with the
same identifier as $t$.
\end{lemma}
{\em Proof}:Suppose $(D,D')\vD m$. By Definition \ref{def:new}, for each
pair of tuples $t_1\in R(D)$ and $t_2\in S(D)$ such that
$t_1[\bar A]\approx t_2[\bar B]$,
$$
t_1'[\bar C] = t_2'[\bar E]\nn
$$
Therefore, if $T(\bar t_1, \bar t_2)$ is true, then
$t_1'$ and $t_2'$ must be in the transitive closure
of the binary relation expressed by
$t_1'[\bar C] = t_2'[\bar E]$. But the
transitive closure of this relation is the
relation itself (because of the transitivity
of equality). Therefore, $t_1'[\bar C] = t_2'[\bar E]$.
The converse is trivial. \boxtheorem

\vspace{2mm}
\defproof{Proposition \ref{lem:MRI}}{Consider an input $D$, $M$ to {\em ComputeMRI}
with $M$ a simple-cycle set of MDs given by
\begin{eqnarray}
R[\bar A_0]\approx_0 S[\bar B_0]&\ra& R[\bar A_0']\rl S[\bar B_0'] \nn\\
R[\bar A_1]\approx_1 S[\bar B_1]&\ra& R[\bar A_1']\rl S[\bar B_1']\nn\\
&\vdots&\nn\\
R[\bar A_{n-1}]\approx_{n-1} S[\bar B_{n-1}]&\ra& R[\bar A_{n-1}']\rl S[\bar B_{n-1}']\nn \label{eq:1}
\end{eqnarray}
Let $T_j$ denote the
transitive closure of the relation $\approx_j$.
Let $D_i$ denote an instance obtained by
updating $D$ $i$ times according to $M$,
and for a tuple $t\in D$, denote the
tuple with the same identifier in $D_i$ by
$t^i$.
By Lemma \ref{lem:tc}
and straightforward induction, it can be
seen that, after $D$ has been updated $i$
times, $i\geq 1$ \footnote{We use the term ``update"
even if a resolved instance is obtained
after fewer than $i$ modifications. In this
case, the ``update" is the identity mapping on all
values.} according to $M$ to
obtain an instance $D_i$, for all tuples $t$ in
a given equivalence class $E$ of $T_j$,
\bea\label{eq:7}
t^i[\bar A_{(j+i-1)\bmod n}'] = \bar v_{ij}^E~~\hbox{ if }t\in R(D)\\
\label{eq:8}
t^i[\bar B_{(j+i-1)\bmod n}'] = \bar v_{ij}^E~~\hbox{ if }t\in S(D)
\eea
for some vector of values $\bar v_{ij}^E$. Let $D'$ be a resolved
instance. $D'$ satisfies
the property that any number of applications of
the MDs does not change the instance. Therefore,
$D'$ must satisfy (\ref{eq:7}) and (\ref{eq:8}) for all $i$.
That is, for any $T_j$, $1\leq j\leq n$, for any equivalence class of
$T_j$, for all tuples $t$ in the
equivalence class, and for $1\leq i\leq n$,
\bea\label{eq:9}
t'[\bar A_i'] = \bar v_{ij}^E~~\hbox{ if }t\in R(D)\\
\label{eq:10}
t'[\bar B_i'] = \bar v_{ij}^E~~\hbox{ if }t\in S(D)
\eea
for some vector of values $\bar v_{ij}^E$, where $t'$ is the tuple in
$D'$ with the same identifier as $t$.

Let $T$ be the transitive closure of the
set $\{T_j|1\leq j\leq n\}$ (cf. definition \ref{def:tcset}).
By (\ref{eq:9}) and (\ref{eq:10}), for any pair of tuples $t_1$ and $t_2$
satisfying $T(t_1,t_2)$, $t_1'$ and $t_2'$ must
satisfy $T'(t_1',t_2')$, where $T'$ is the transitive
closure of the binary relation on tuples
expressed by $t_1'[\bar A_i'] = t_2'[\bar B_i']$, $1\leq i\leq n$.
Since the equality relation is closed under
transitive closure, this implies the following property:
\bea\label{eq:11}
T(t_1,t_2) \hbox{ implies } t_1'[\bar A_i'] = t_2'[\bar B_i'],~ 1\leq i\leq n
\eea
It remains to show that the instances produced
by {\em ComputeMRI} are resolved instances. That they
are the MRIs will then follow from the fact that they
have the fewest changes among all instances satisfying
(\ref{eq:11}). For any equivalence class $E$ of
$T$, let $\bar v_i^E$ be a list of values chosen by {\em ComputeMRI} as
the common values for the pair of attribute lists $(\bar A_i',\bar B_i')$
for tuples in $E$. To obtain the instance output
by {\em ComputeMRI} for this choice of values, $D$ can be updated
as follows.
For the $i^{th}$ update, if the values of the
attributes $\bar A_i'$ and $\bar B_i'$ must be modified to achieve (\ref{eq:7})
and (\ref{eq:8}), take $\bar v^E_{ij} = \bar v_i^{E'}$, where $E'$ is the equivalence
class of $T$ that contains the equivalence class $E$ of $T_j$.
Note that such an $E'$ always exists, and the
assignment of values is consistent since overlapping
equivalence classes $T_i$ and $T_j$ will be contained in
the same equivalence class of $T$. Then after
$n$ updates, the resulting instance satisfies (\ref{eq:11}), with
common values as chosen by {\em ComputeMRI}.

We must show that the resolved instance produced by
this update process is the same instance that {\em ComputeMRI}
returns for the given choice of update values.
For any intermediate instance $I$ obtained in this
update process, let $t_I$ denote the tuple in $I$
with the same identifier as $t$. We will show by
induction on the number of updates that were made
to obtain $I$ that for any $i$, whenever $T_i(t_I,t_I')$
for tuples $t$ and $t'$, it holds that $T(t,t')$. This
implies that updates made to $t[A]$ for tuple $t$
and attribute $A$ can only set it equal to the common
value for the equivalence class of $T$ to which $t$
belongs. Since {\em ComputeMRI} also sets $t[A]$ to this
value, this will prove the theorem.

By definition of $T$, if 0 updates were used to
obtain $I$, $T_i(t_I,t_I')$ implies $T_i(t,t')$ implies
$T(t,t')$. Assume it is true for instances obtained
after at most $k$ updates. Let $I$ be an instance
obtained after $k+1$ updates. Suppose for the
sake of contradiction that there exist tuples
$t_I$ and $t_I'$ such that for some $i$,
$T_i(t_I,t_I')$ but $\lnot T(t,t')$.
Since $\lnot T(t,t')$ implies $\lnot T_i(t,t')$,
at least one of $t[\bar A_i']$ and $t'[\bar B_i']$ was
updated so that $T_i(t_I,t_I')$. We will assume that
only $t[\bar A_i']$ was updated.
The other cases are similar. Then
it must have been updated to $t''[\bar A_i']$
or $t''[\bar B_i']$ for some $t''\in R$ or
$t''\in S$, respectively, such that,
for the instance $I'$ on which the
update was performed, it holds that
$T_{i}(t_{I'},t_{I'}'')$ and $T_{i}(t_{I'}',t_{I'}'')$. By the
induction hypothesis, $T(t,t'')$ and $T(t',t'')$,
which by the transitivity of $T$ implies $T(t,t')$,
a contradiction. }

\vspace{2mm}
\defproof{Theorem \ref{theo:hsc}}{
If it can be verified in
polynomial time that an instance is an MRI of a given
instance wrt a set $M$ of MDs, then $RA_{Q,M}$ is
in co-NP for any FO $Q$. This is because, for a given instance $(D,t)$
of $RA_{Q,M}$, $t$ can be
shown not to be a certain answer by guessing an
instance $D'$, verifying that it is an MRI,
and verifying that $t$ is not an answer to $Q$
for $D'$. Algorithm {\em ComputeMRI} can
easily be modified to produce such a polynomial
time verifier: compute the transitive closure
relation $T$ but instead of setting values
equal, check that they are equal in the
candidate MRI.}


\begin{lemma}\label{lem:onestep} \
Let $M$ be a non-interacting set of MDs of the form
\bea \label{eq:2n}
m_1:~~&&R_1[A_1]\approx_1 R_2[B_1]\ra R_1[\bar A_2]\rl R_2[\bar B_2]\nn\\
m_2:~~&&R_3[A_3]\approx_2 R_4[B_3]\ra R_3[\bar A_4]\rl R_4[\bar B_4]\nn\\
&&\vdots\nn\\
m_{n}:~~&&R_{2n-1}[A_{2n-1}]\approx_n R_{2n}[B_{2n-1}]\ra\nn\\
&& R_{2n-1}[\bar A_{2n}]\rl R_{2n}[\bar B_{2n}]\nn
\eea
Let $T$ be the transitive closure of the set $\{TA_1,TA_2,...TA_n\}$,
where $TA_i$ is the tuple-attribute closure of $m_i$. Then, for
any instance $D$, an instance $D'$ obtained by updating
modifiable values of $D$ is a resolved instance of $D$
iff whenever $T(t_1,A,t_2,B)$, $t_1'[A] = t_2'[B]$,
where $t'$ is the tuple in $D'$ with the same identifier
as $t$.
\end{lemma}
{\em Proof}: Suppose $D'$ is a resolved instance.
Since $M$ is non-interacting, this implies $(D,D')\vD M$.
It is a corollary of Lemma \ref{lem:tc}
that whenever $T(t_1,A,t_2,B)$, $t_1'[A] = t_2'[B]$,
for all $1\leq i\leq n$. The converse
follows from the fact that, whenever a pair of tuples
$t_1$ and $t_2$ satisfies the similarity condition of
an MD, $T(t_1,A,t_2,B)$ for every pair $(A,B)$ of matched
attributes in the MD. \boxtheorem

\vspace{2mm}
\begin{corollary}\label{cor:MRIs}
Let $D$ be an instance and $M$ a set of non-interacting MDs.
Let $T$ be the transitive closure of the
set of tuple-attribute closures of the MDs
in $M$. Then the set of MRIs is obtained
by setting, for each equivalence class $E$ of
$T$, the value of each attribute in $E$ to a
value that occurs in $E$ at least as frequently
as any other value in $E$.
\boxtheorem
\end{corollary}

\vspace{2mm}
\defproof{Theorem \ref{thm:rewrite}}{ We express the query in the
form
\bea\label{eq:form}
\mathcal{Q}(\bar y) = \ex\bar z Q_1(\bar z,\bar y)
\eea
Let $x_{ij}$ denote the variable of $\bar z$
or $\bar y$ which holds the value of the $j^{th}$
attribute in the $i^{th}$ conjunct $R_i$ in $Q_1$.
Denote this attribute by $A_{ij}$.
Note that, since variables and conjuncts can be repeated,
it can happen that $x_{ij}$ is the same variable
as $x_{kl}$ for $(i,j)\neq (k,l)$, that $A_{ij}$ is the same
attribute as $A_{kl}$ for $(i,j)\neq (k,l)$, or
that $R_i$ is the same as $R_j$ for $i\neq j$.
Let $B$ and $F$ denote the set of bound
and free variables in $Q_1$, respectively.
Let $C$ and $U$ denote the variables in $Q_1$ holding
the values of changeable and unchangeable attributes,
respectively.
Let $\mathcal{Q}'(\bar y)$ denote
the rewritten query returned by algorithm {\em Rewrite}, which
we express as
$$
\mathcal{Q}'(\bar y) = \ex z Q_1'(\bar z,\bar y)
$$
We show that, for any constant vector $\bar a$, $\mathcal{Q}'(\bar a)$
is true for an instance $D$ iff $\mathcal{Q}(\bar a)$ is true
for all MRIs of $D$.

Suppose that $\mathcal{Q}'(\bar a)$ is true for an instance $D$.
Then there exists a $\bar b$ such that $Q_1'(\bar a,\bar b)$.
We will refer to this assignment of constants to variables
as $A_{\mathcal{Q}'}$.
From the form of $\mathcal{Q}'$, it is apparent that,
for any fixed $i$, there is a tuple
$t_1 = \bar c_i \equiv (c_{i1},c_{i2},...c_{ip})$ such that
$R_i(\bar c_i)$ is true in $D$ with the following properties.
\begin{enumerate}
\label{enum:one}
\item For all $x_{ij}$ except those in $F\bc C$, $c_{ij}$ is
the value assigned to $x_{ij}$ by $A_{\mathcal{Q}'}$.
\label{enum:two}
\item For all $x_{ij}\in F\bc C$, there is a tuple $t_2$
with attribute $B$ such that $T(t_1,A_{ij},t_2,B)$,
where $T$ is the transitive closure of the tuple-attribute
closures of the MDs in $M$, such that
the value of $t_2[B]$ is the value assigned to $x_{ij}$
by $A_{\mathcal{Q}'}$.
Moreover, this value
occurs more frequently than that
of any other tuple/attribute pair in the same equivalence
class of $T$.
\end{enumerate}
For any given MRI $D'$, consider the tuple $t_1'$ in $D'$ with the
same identifier as $t_1$. Clearly, this tuple will have the
same values as $t_1$ for all unchangeable attributes, which
by 1., are the values assigned to the variables
$x_{ij}\in U$. Also, by 2.
and Corollary \ref{cor:MRIs}, for any $j$ such that $x_{ij}\in F\bc C$
is free, the value of the $j^{th}$ attribute of $t_1'$ is
that assigned to $x_{ij}$ by $A_{\mathcal{Q}'}$.

Thus, for each MRI $D'$, there exists an assignment
$A_{\mathcal{Q}}$ of constants to the $x_{ij}$ that makes $\mathcal{Q}$ true, and
this assignment agrees with $A_{\mathcal{Q}'}$ on all $x_{ij}\nin B\bc C$.
This assignment is consistent in the
sense that, if $x_{ij}$ and $x_{kl}$ are the same variable,
they are assigned the same value. Indeed, for $x_{ij}\nin B\bc C$,
consistency follows from the consistency of $A_{\mathcal{Q}'}$, and for
$x_{ij}\in B\bc C$, it follows from the fact that the variable
represented by $x_{ij}$ occurs only once in $Q$, by assumption.
Therefore, $\mathcal{Q}(\bar a)$ is true for all MRIs $D'$, and $\bar a$
is a resolved answer.

Conversely, suppose that a tuple $\bar a$ is a resolved
answer. Then, for any given MRI $D'$ there is a satisfying
assignment $A_\mathcal{Q}$ to the variables in $\mathcal{Q}$ such that
$\bar z$ as defined by (\ref{eq:form}) is assigned the
value $\bar a$. We write $\mathcal{Q}'$ in the form
\bea\label{eq:form2}
\mathcal{Q}'(\bar y)\lar \ex \bar z\w_{1\leq i\leq n}Q_i(\bar v_i)
\eea
with $Q_i$ the rewritten form of the $i^{th}$ conjunct
of $\mathcal{Q}$. For any fixed $i$, let $t' = (c_{i1}',c_{i2}',...c_{ip}')$ be a tuple
in $D'$ such that $c_{ij}'$ is the constant assigned to
$x_{ij}$ by $A_\mathcal{Q}$.

We construct a satisfying assignment $A_{\mathcal{Q}'}$ to the free
and existentially quantified variables of $\mathcal{Q}'$ as
follows. Consider the conjunct $Q_i$ of $\mathcal{Q}'$
as given on line 16 of {\em Rewrite}.
Assign to $\bar v_i'$ the tuple $t$
in $D$ with the same identifier as $t'$. This fixes the
values of all the variables except those $x_{ij} \in F\bc C$,
which are set to $c_{ij}'$. It follows from Corollary \ref{cor:MRIs}
that $A_{\mathcal{Q}'}$ satisfies $\mathcal{Q}'$. Since $A_{\mathcal{Q}}$ and $A_{\mathcal{Q}'}$ match
on all variables that are not local to a single $Q_i$,
$A_{\mathcal{Q}'}$ is consistent. Therefore, $\bar a$ is an
answer for $\mathcal{Q}'$ on $D$.}

\vspace{2mm}

\defproof{Theorem \ref{thm:Chom}}{Hardness follows from the
fact that, for the instance $D$ resulting from the
reduction in the proof of Theorem 3.3 in \cite{Chomicki05}, the set
of all repairs of $D$ with respect to the given key constraint
is the same as the set of MRIs with respect to
(\ref{eq:intMD}). The key point is that
attribute modification in this case
generates duplicates which are subsequently eliminated
from the instance, producing the same
result as tuple deletion. Containment follows from
Theorem \ref{theo:hsc}.}

\vspace{2mm}
\defproof{Proposition \ref{thm:red}}{Take $\bar A = (A_1,...A_m)$
and $\bar B = (B_1,...,$ $B_n)$. For any tuple
of constants $\bar k$, define
$R^{\bar k}\equiv\s_{\bar A = \bar k}R$.
Let $B_i^{\bar k}$ denote the single attribute
relation with attribute $B_i$ whose tuples are
the most frequently occurring values in
$\pi_{B_i}R^{\bar k}$. That is, $a\in B_i^{\bar k}$
iff $a\in \pi_{B_i}R^{\bar k}$ and there is no
$b\in \pi_{B_i}R^{\bar k}$ such that $b$ occurs
as the value of the $B_i$ attribute in more tuples
of $R^{\bar k}$ than $a$ does. Note that
$B_i^{\bar k}$ can be written as an
expression involving $R$ which is first order
with a $\nit{Count}$ operator.
The reduction produces $(R',t)$ from $(R,t)$,
where
\bea
R'\equiv \bigcup_{\bar k} \l[\pi_{\bar A}R^{\bar k}
\times B_1^{\bar k}\times\cdots B_n^{\bar k}\r]
\eea
The repairs of $R'$ are obtained by keeping,
for each set of tuples with the same key
value, a single tuple with that key value and
discarding all others. By Corollary A.\ref{cor:MRIs},
in a MRI of $D$, the group $G_{\bar k}$ of tuples such that
$\bar A = \bar k$ for some constant $\bar k$
has a common value for $\bar B$ also,
and the set of possible values for $\bar B$
is the same as that of the tuple with key $\bar k$
in a repair of $D$. Since duplicates are eliminated
from the MRIs, the set of MRIs of $D$ is exactly the
set of repairs of $R'$.}

\vspace{2mm}
\defproof{Theorem \ref{theo:comp}}{$\mathcal{Q}''$ is obtained by composing
$\mathcal{Q}'$ with the transformation $R\ra R'$, which is
a first-order query with aggregation.}

\end{document}